MDPI

*Article*

# Multimodal Approaches for Indoor Localization for Ambient Assisted Living in Smart Homes

**Nirmalya Thakur** * 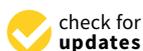 **and Chia Y. Han**

Department of Electrical Engineering and Computer Science, University of Cincinnati,
Cincinnati, OH 45221-0030, USA; han@ucmail.uc.edu
* Correspondence: thakurna@mail.uc.edu

**Abstract:** This work makes multiple scientific contributions to the field of Indoor Localization for Ambient Assisted Living in Smart Homes. First, it presents a Big-Data driven methodology that studies the multimodal components of user interactions and analyzes the data from Bluetooth Low Energy (BLE) beacons and BLE scanners to detect a user's indoor location in a specific 'activity-based zone' during Activities of Daily Living. Second, it introduces a context independent approach that can interpret the accelerometer and gyroscope data from diverse behavioral patterns to detect the 'zone-based' indoor location of a user in any Internet of Things (IoT)-based environment. These two approaches achieved performance accuracies of 81.36% and 81.13%, respectively, when tested on a dataset. Third, it presents a methodology to detect the spatial coordinates of a user's indoor position that outperforms all similar works in this field, as per the associated root mean squared error—one of the performance evaluation metrics in ISO/IEC18305:2016—an international standard for testing Localization and Tracking Systems. Finally, it presents a comprehensive comparative study that includes Random Forest, Artificial Neural Network, Decision Tree, Support Vector Machine, k-NN, Gradient Boosted Trees, Deep Learning, and Linear Regression, to address the challenge of identifying the optimal machine learning approach for Indoor Localization.

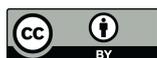

## 1. Introduction

Technologies like Global Positioning Systems (GPS) and Global Navigation Satellite Systems (GNSS) have revolutionized navigation research by being able to track people, objects, and assets in real-time. Despite the significant success of these technologies in outdoor environments, they are still ineffective in indoor settings [1]. This is primarily for two reasons, first, these technologies depend on line of sight communication between GPS satellites and receivers which is not possible in an indoor environment and second, GPS provides a maximum accuracy of up to five meters [2]. With Industry 4.0, there has been an increasing need for developing systems for indoor navigation and localization for the future of living and working environments, which would involve human–computer, human–machine, and human–robot interactions in a myriad of ways. These environments could involve Smart Homes, Smart Cities, Smart Workplaces, Smart Industries, and Smart Vehicles, just to name a few. There are multiple application domains that are in need for a standard methodology for Indoor Localization. A system for Indoor Localization may broadly be defined as a system of interconnected devices, networks, and technologies that help to detect, track, and locate the position of people and objects inside closed or semi-closed environments, where technologies such as GPS or GNSS do not work [3]. As per [3], the market opportunities of Indoor Localization related systems are expected to be in the order of USD 10 billion by 2024 due to the diverse societal needs that such systems can address. Some potential applications of such Indoor Localization related technologies could include (1) tracking the location of products during smart manufacturing in automated or semi-automated manufacturing sites; (2) tracking the location and operation of unmanned vehicles or robots in industrial settings; (3) detecting the precise location of an elderly fall for communicating the same to emergency responders; (4) helping older adults with various forms of Cognitive Impairments (CI) to perform their daily routine tasks by directing them to specific locations for performing these activities; (5) tracking the precise location of individuals with Dementia or Alzheimer's when they face freezing of gait to alert caregivers; (6) assisting the visually impaired to reach specific objects of interest in both living and working environments; (7) helping individuals suffering from delirium to navigate from one place to the other for performing different activities; (8) detection of the precise location of the elderly when they face cramps or other forms of motor impairments; (9) detecting the location of patients in hospitals to avoid the need for in-person monitoring; and (10) automated tracking of different kinds of physical assets in Internet of Things (IoT)-based functional and work-related environments.

Only one area of interest, Ambient Assisted Living (AAL) of Elderly People during Activities of Daily Living (ADLs) in the future of technology-laden living environments, for instance, Smart Homes and Smart Cities, will be addressed in this work. AAL may broadly be defined as a computing paradigm that uses information technology and its applications to enhance user abilities, performance, and quality of life through interconnected systems that can sense, anticipate, adapt, predict, and respond to human behavior and needs. Human behavior in the confines of their living and functional environments is characterized by activities that they perform in these environments. In a broad scope, an activity may be defined as an interaction between a subject and an object, for the subject to achieve a desired end goal or objective. Here, the subject is the user who performs the activity and the set of environment parameters that they interface with during this activity are known as the objects. Based on the variations in the environment in which the activity is performed, the same activity may involve different objects that a user interfaces with, to reach the end goal. Similarly, the diversities in the user can also lead to different interaction patterns with objects for performing the same activity in the same or in a different environment [4]. Activities can have various characteristic features. These include—sequential, concurrent, interleaved, false start, and social interactions. Those activities that are crucial for one's sustenance and which one performs on a daily routine basis are known as Activities of Daily Living (ADLs). There are five broad categories of ADLs—Personal Hygiene, Dressing, Eating, Maintaining Continence, and Mobility [5].

People live longer these days due to advanced healthcare facilities. The population of elderly people has been on a constant rise and there are around 962 million elderly people [6] across the world. According to [7], by 2050, the population of elderly people is expected to become around 1.6 billion and outnumber the population of younger people globally. To add, the population of older adults, aged 80 years or more, is expected to increase three times and reach around 425 million by 2050. Increasing age is associated with physical disabilities, cognitive impairments, memory issues, and disorganized behavior which limit a person's ability to carry out their daily routine tasks in an independent manner. The worldwide costs of looking after elderly people with various forms of cognitive impairments, such as Dementia, is estimated to be around USD 818 billion and is increasing at a very fast rate [8]. In the United States alone, approximately 5.8 million elderly people currently have Dementia and 1 in every 3 seniors dies from Dementia. In 2020, care of people with Dementia accounted for approximately USD 305 billion to the U.S. economy, out of which the caregiver costs are estimated to be around USD 244 billion. It is predicted that these costs are going to rise steeply over the next few years [8]. A major challenge in this field is to make the future of Internet of Things (IoT)-based ubiquitous living environments, such as Smart Homes and Smart Cities aware, adaptive, and personalized so that they can contribute towards independent living and healthy aging of the elderly while fostering their biological, psychological, behavioral, physical, mental, and emotional well-being. Indoor Localization has an immense role to play towards addressing these

challenges—both in terms of independent living and healthy aging of the elderly as well as for addressing the huge burden of their caregiving costs.

Despite several advances [9–59] in this field, which have been reviewed in a detailed manner in Section 2, multiple research challenges remain to be addressed. These include—(1) inability of the activity recognition and the activity analysis-based AAL systems to track the indoor location of the user during ADLs; (2) dependency of Indoor Localization systems on context parameters local to specific IoT-based settings which limit the functionalities of such systems to those specific environments; (3) need for better precision and accuracy for detection of the indoor location of a user; and (4) need to deduce and identify the optimal machine learning-based approach in view of the wide variety of learning approaches that have been investigated by researchers for development of Indoor Localization systems. Thus, addressing these above-mentioned challenges by exploring the intersections of Big Data, Machine Learning, Indoor Localization, Ambient Assisted Living, Internet of Things, Activity Centric Computing, Human–Computer Interaction, Pattern Recognition, and Assisted Living Technologies to provide a long-term, robust, feasible, easily implementable, sustainable, and economic solution to these global research challenges serves as the main motivation for the work presented in this paper. To summarize, the scientific contributions of this paper are as follows:

1. Big-Data driven methodology that studies the multimodal components of user interactions and analyzes the data from BLE beacons and BLE scanners to track a user's indoor location in a specific 'activity-based zone' during Activities of Daily Living. This approach was developed by using a k-nearest neighbor (k-NN)-based learning approach. When tested on a dataset it achieved a performance accuracy of 81.36%.

2. A context independent approach that can interpret the accelerometer and gyroscope data from diverse behavioral patterns to detect the 'zone-based' indoor location of a user in any IoT-based environment. Here, the 'zone-based' mapping of a user's location refers to mapping the user in one of the multiple 'activity-based zones' that any given IoT-based environment can be classified into based on the associated context attributes. This methodology was developed by using a Random Forest-based learning approach. When tested on a dataset it achieved a performance accuracy of 81.13%.

3. A methodology to detect the spatial coordinates of a user's indoor position based on the associated user interactions with the context parameters and the user-centered local spatial context, by using a reference system. The performance characteristics of this system were evaluated as per three metrics stated in ISO/IEC18305:2016 [31], which is an international standard for testing Localization and Tracking Systems. These metrics included root mean squared error (RMSE) in X-direction, RMSE in Y-direction, and the Horizontal Error, which were found to be 5.85 cm, 5.36 cm, and 7.93 cm, respectively. A comparison of the performance characteristics of this approach with similar works in this field that used the RMSE evaluation method showed that our system outperformed all recent works that had a similar approach.

4. A comprehensive comparative study of different machine learning approaches that include—Random Forest, Artificial Neural Network, Decision Tree, Support Vector Machine, k-NN, Gradient Boosted Trees, Deep Learning, and Linear Regression, with an aim to address the research challenge of identifying the optimal machine learning-based approach for Indoor Localization. The performance characteristics of each of these learning methods were studied by evaluating the RMSE in X-direction, the RMSE in Y-direction, and the Horizontal Error as per ISO/IEC18305:2016 [31]. The results and findings of this study show that the Random Forest-based learning approach can be considered as the optimal learning method for development of Indoor Localization and tracking related technologies.

This paper is organized as follows. We present a comprehensive overview of the related works in this field in Section 2. In Section 3, a brief overview is given about RapidMiner [60], a data science and machine learning software development platform, that has been used for development of all the methodologies proposed in this paper. Section 4 presents the methods and the steps associated with the development of the three novel methodologies for Indoor Localization that have been proposed in this work. Section 5 discusses the results and findings associated with each of these methodologies. In Section 6, we present the comparative study of different machine learning approaches that include—Random Forest, Artificial Neural Network, Decision Tree, Support Vector Machine, k-NN, Gradient Boosted Trees, Deep Learning, and Linear Regression, with an aim to address the research challenge of identifying the optimal machine learning-based approach for Indoor Localization. Section 7 elaborates the research challenges in this field and discusses how the work presented in this paper outperforms all similar works while addressing the associated research challenges. It is followed by Section 8 where conclusion and scope for future work are outlined.

## 2. Literature Review

In this section, we have reviewed different kinds of AAL-based systems and technologies that have primarily focused on Indoor Localization, Activity Recognition, and Activity Analysis in Smart and Interconnected IoT-based environments, such as Smart Homes and Smart Cities.

Machine learning approaches have been widely used by researchers to track people and objects in indoor environments and settings. Musa et al. [9] developed a system that used a non-line of sight approach and multipath propagation in the context of using the ultra-wide band methodology. The system used a cross-fold validation method to train a decision tree that could detect the indoor location of a user. A similar decision-tree driven machine learning framework was developed by Yim et al. [10]. The framework was equipped with the functionality to build the decision tree in the off-line phase and it used the fingerprinting approach for Indoor Localization. Sjoberg et al. [11] developed a visual recognition approach using the support vector machine (SVM) classifier. The system consisted of a visual bag-of-words model and other visual features of the environment that were used to train this classifier for Indoor Localization. A method of 2.5D indoor positioning was proposed by Zhang et al. [12]. Here, the SVM classifier was trained to detect the specific floor where a user is located based on the WiFi signal strength and thereafter it obtained the user's position information by analysis of other characteristics of the associated altitude data. Zhang et al. [13] developed a k-NN classification approach for Indoor Localization that used the signal strength fingerprint technology. The system assigned weights to the samples based on their associated signal strengths to divide them into clusters, where each cluster represented a specific location.

In [14], Ge et al. proposed an algorithm for indoor location tracking that was developed by using the k-NN approach. The algorithm used signal processing principles to detect and analyze the data coming from access points in the user's location to train the k-NN classifier. In [15], Hu et al. proposed another k-NN based learning approach that detected the location of the user based on the nearest access points of the user. One of the key findings of the work was that the condition of k = 1 led to the best positioning performance accuracy. The artificial neural network approach (ANN) was used by Khan et al. [16] for developing an indoor position detection system. The architecture of the approach involved studying and interpreting the data from Wireless Local Area Network (WLAN) access points and Wireless Sensor Networks (WSN) to train the artificial neural network (ANN) that could perform virtual tessellation of the available indoor space. Another neural network driven system was proposed by Labinghisa et al. [17]. This approach was based on the concept of virtual access points with an aim to increase the number of access points without the requirements of any additional hardware. These additional access points helped to track more user movement related data for training of the neural network.

A Wi-Fi fingerprint-based indoor positioning system was proposed by Qin et al. [18] that was neural network driven. The system used a convolutional denoising autoencoder to analyze and extract key features from the RSSI data, which were then used to train the neural network. The authors evaluated their system on two datasets to discuss its performance characteristics. In [19], Varma et al. used the Random Forest learning approach to perform Indoor Localization in an Internet of Things (IoT)-based environment during real-time experiments. The authors set up an IoT-based space with 13 beacons. The signals coming from these beacons, based on the user's varying position, were used to train the Random Forest model. A Wi-Fi signal analysis based indoor position detection system was proposed by Gao et al. [20] that also used the Random Forest classifier. It was developed and implemented by using the region-based division of location grids method to minimize the maximum error. The system adopted the method of adjusted cosine similarity to match the user's position with the exact grid by analyzing the fingerprint information. Linear regression methods have also been used by researchers to develop Indoor Localization systems and technologies. For instance, in [21], the authors developed a learning model where each anchor node had its own linear ranging method. Their linear regression model studied the distance between different anchors to perform anchor distance-based location detection.

The work proposed by Barsocchi et al. [22] involved development of a linear regression-based learning approach that calculated the user's distance from a reference point based on RSSI values. The approach consisted of the methodology to map these values into distances to detect the position of the user in the given environment. Zhang et al. [23] developed a deep learning-based 3D positioning framework for a hospital environment. The system used the data from the cell phone network as well as the Wi-Fi access points to determine the exact position of the user in terms of the latitude, longitude, and the level of the building at which they were present. Another deep learning-based Indoor Localization system was proposed by Poulose et al. [24]. The system developed heat maps from the RSSI signals obtained from the access points, to train the deep learning model. By conducting experiments, the authors evaluated the effectiveness of their system for its deployment in an autonomous environment. A Gradient Boosted Decision Tree approach was proposed by Wang et al. [25] that used the fingerprint methodology to detect the location of a user in an indoor setting. The authors used the concept of wavelet transform to filter the noises in the channel state information data, which was then used by the system to update the associated fingerprint information for the machine learning model. As can be seen from [9–25] a range of machine learning approaches have been used for development of various types of Indoor Localization systems for IoT-based environments. However, none of these works implemented multiple machine learning models to evaluate and compare their working to deduce the best machine learning approach in terms of the associated performance accuracy. Due to the differences in the datasets used or the real-time data that was collected, the associated data preprocessing steps that varied from system to system, the differences in train to test ratio of the data, and several other dissimilar steps that were associated with the developments of each of these machine learning models, their final performance accuracies cannot be directly compared to deduce the best approach. The challenge is therefore to identify the optimal machine learning model that can be used to develop the future of Indoor Localization systems, Indoor Positioning systems, and Location-Based Services.

In addition to machine learning-based approaches, context reasoning-based approaches have also been investigated by researchers to develop Indoor Localization systems. Such approaches are limited and functional only in the confines of the specific environments for which they were developed. In [26], Lin et al. developed an indoor positioning system specific for the factory environments of the Hon-Hai Precision Industry. An intelligent indoor parking system was developed by Liu et al. [27] that could help with indoor parking. The system was implemented in a shopping mall environment to test its performance characteristics. Jiang et al. [28] proposed an Indoor Localization system for hospital environments that used concepts from GPS and UWB technologies. Barral et al. [29] developed a methodology that could track the location of forklift trucks in an industry-based environment. Zadeh et al. [30] proposed an Indoor Localization framework for an academic environment to help with taking attendance of students. As can be seen from [26–30], these Indoor Localization systems are specific to certain environments, as they are dependent on the associated features and characteristics of the environment for which they were designed. For instance, the system proposed in [29] cannot be deployed in any of the environments described in [26–28,30]. The environments described in [26–30] represent just a few of the IoT-based environments associated with the living and functional areas of humans in the future of interconnected Smart Cities. As such context reasoning-based systems are not functional in other IoT-based settings, the challenge is thus to develop a means for Indoor Localization that is not environment dependent and can be seamlessly deployed in any IoT-based setting irrespective of the associated context parameters and their attributes.

Here, we also review ISO/IEC 18305:2016, which is an international standard for evaluating localization and tracking systems [31]. This standard was jointly prepared and developed by the Joint Technical Committee ISO/IEC JTC 1, Information technology, Subcommittee SC 31. ISO/IEC JTC 1/SC 31 is a standardization subcommittee of the joint committee ISO/IEC JTC 1 of the International Organization for Standardization (ISO) and the International Electrotechnical Commission (IEC), which develops and facilitates international standards, technical reports, and technical specifications in the field of automatic identification and data capture techniques. This standard is one of the outcomes of the EU FP7 project EVARILOS—Evaluation of RF-based Indoor Localization Solutions for the Future Internet [32]. The main objective of this standard is to define a standard set of testing and evaluation measures or methods that can be used to evaluate the performance metrics of different types of Indoor Localization systems, Indoor Positioning systems, and Location Based Services in different scenarios. It provides a comprehensive list of 14 such scenarios and 5 types of buildings where this standard can be implemented. It is worth mentioning here that the standard discusses these settings from the viewpoint of localization of a person, object as well as a robot in such scenarios. Such definitions of scenarios include the characterization of the associated motion as well as the definition of the number of entities (human, objects, or robots) that are required to be tracked in that given scenario. It also lists 30 metrics for evaluating the performance characteristics in each case. The metrics related to calculation of different kinds of errors are introduced in Chapter 8 of this standard. Some of these metrics that have been widely used by researchers since the inception of this standard include—RMSE, Mean of Error, Covariance Matrix of Error, Mean of Absolute Error, and Mean and Standard Deviation of Vertical Error.

While several of these metrics have been used by researchers to evaluate their Indoor Localization systems, we focus on one specific performance metric—the RMSE. The standard presents the formulae for determination of the RMSE in the X-direction, RMSE in the Y-direction, and RMSE in the X-Y plane. When the RMSE is determined in the X-Y plane, it is referred to as Horizontal Error as per the definitions of the standard [31]. Next, we review some of the works related to Indoor Localization systems that have used the RMSE method for evaluating their performance characteristics. In [33], the authors analyzed the RSSI data coming from multiple anchor nodes set up in a Wireless Sensor Network system. They used Kalman filter to determine the direction and speed of the user and their system had a RMSE of 1.4 m. In [34], the authors developed a new RFID-based device that could sense proximity tags in the environment to detect the indoor location of a user with a RMSE of 0.32 m. Angermann et al. [35] developed a Bayesian estimation-based framework for pedestrian localization and mapping that had a RMSE of 1 to 2 m. Evennou et al. [36] used signal processing-based methods to develop an Indoor Localization system that had a RMSE of 1.53 m. Wang et al. [37] used particle filters and extended the traditional WLAN methods to develop a pedestrian tracking system that had a RMSE of 4.3 m. A Monte Carlo-based Indoor Localization algorithm was proposed in [38]. The RMSE of this system was 1.2 m. A SVM classifier was proposed in [39] that used smartphone data for performing Indoor Localization. The performance characteristics presented in the paper show that the value of the RMSE of this system was 4.55 m. Another smart phone-based system, known as HIVE [40], that was Hidden Markov Model driven was proposed by Liu et al.

The system had a RMSE of 3.1 m. The work by Chen et al. [41] involved fusion of data coming from smartphone sensors, WiFi, and Landmarks, which were analyzed by a Kalman Filter for Indoor Localization. This method had a RMSE of 1 m. Li et al. [42] proposed a sensor technology-driven smartphone-based pedestrian location detection system that had a RMSE of 2.9 m. In [43], the authors analyzed iBeacon measurements and a calibration range, which was thereafter used to develop a Kalman filter for pedestrian dead reckoning. The system had a RMSE of 1.28 m. As per [44] and [45], (1) the average dimensions of newly built one-bedroom apartments and two-bedroom apartments in United States in 2018 were 757 square feet (70.3276 square meters) and 1138 square feet (105.7236 square meters), respectively. Considering such dimensions of living spaces in the context of AAL in Smart Homes, it is the need of the hour that the Indoor Localization systems and tracking related technologies become more precise in terms of detecting the exact indoor location of the user. The challenge is this context is thus to develop Indoor Localization systems that have lesser RMSE as compared to [33–43].

AAL in Smart Homes is not only about tracking the indoor location of the user, it also involves analyzing their behaviors and activity patterns to enhance their quality of life and user experience in the context of diverse user interactions. Next, we review some of the recent AAL-based technologies that have focused on activity recognition and analysis in Smart Homes. An activity recognition framework was proposed by Ranieri et al. [46] for AAL of elderly in Smart Homes. The framework focused on studying various parameters of user interaction data from videos, wearable sensors, and ambient sensors to analyze activities. An activity analysis approach for monitoring elderly behavior during daily activities was proposed by Fahad et al. [47]. The objective of the work was to track elderly behavior and detect any possible anomalies in the same that could have resulted from cognitive or physical impairments or decline in abilities. A smart phone accelerometer-based activity recognition framework was proposed by Suriani et al. [48]. The authors developed the system by using spiking neural networks and used datasets to evaluate its performance characteristics. A similar smartphone-based application was proposed by Mousavi et al. in [49] which could detect falls in elderly. The system was trained using SVM and had a performance accuracy of 96.33%. A wearable sensor driven fall detection system was proposed by Alarifi [50]. The wearables were placed on six different locations on the user's body to track multimodal components of motion and behavior data, which were studied and analyzed by a convolution neural network. In [51], Al-Okby proposed a smart wearable device for detection of elderly falls. The device analyzed multimodal components of the user's motion and had the functionality to alert caregivers in the event of a fall. As can be seen from these works [46–51], multiple components of human postures, pose, motion, and behavior can be tracked and analyzed for development of AAL-based activity recognition and fall detection systems. However, the main limitation of these systems is their inability to track the location of the user. For instance, consider the example of an elderly staying alone in an apartment located in a multistoried building. When this elderly person experiences a fall, a fall detection system such as [51] could alert caregivers but the lack of the precise location information could cause delay of medical attention or assistive care. This is because the location of the elderly can be tracked in terms of the building information from GPS, but the specific floor, apartment, or room-related information is not available to the caregivers or emergency responders. Such delay of care can have both short-term and long-term health-related impacts to the elderly. Thus, it is the need of the hour that AAL-based systems not only track, monitor, and analyze elderly behavior but they should also be equipped with the functionality to detect the indoor location of the elderly.

Cloud computing-based approaches have also been used in recent studies [52–55] for development of AAL-based systems and applications that can monitor human behavior and trigger alarms as well as track the location of the user. While this concept of cloud computing applied to AAL technologies holds potential, but these existing systems also have several limitations. For instance—the system proposed by Navarro et al. [53] is environment specific and was designed, adapted, and built specifically for Fundació Ave Maria [56], which is a non-profit organization in Spain, so, the same design cannot not be seamlessly applied to any other environment consisting of varying environment parameters that would be associated with diverse range of human behavior and user interactions; the system proposed by Nikoloudakis et al. [52] uses an outdoor positioning mechanism that can only detect whether the user leaves the premises of their location; the work proposed by Facchinetti et al. [55] is a mobile app and that brings into context these challenges—(i) elderly people are less likely to download a mobile app as compared to the other age groups [57], (ii) elderly people are naturally resistant to using different kinds of technology-based apps on their phones, tablets, and other interactive devices [58], and (iii) older adults face multiple usability issues with such apps [59]; even though another system proposed by Nikoloudakis et al. [54] presents approaches for both indoor and outdoor positioning, it cannot model and analyze the fine grain levels of human activity such as atomic activity, context attributes, core atomic activity, and core context attributes along with their associated weights, in the context of dynamic user interactions during ADLs. To add to the above, for all these systems [52–55], the RMSE for detection of the indoor location of the user is also not less than 1 m. Thus, to summarize, the main research challenges in this field are as follows:

1.  The AAL-based activity recognition, activity analysis, and fall detection systems currently lack the ability to track the indoor location of the user. It is highly essential that in addition to being able to track, analyze, and interpret human behavior, such systems are also able to detect the associated indoor location information, so that the same can be communicated to caregivers or emergency responders, to facilitate a timely care in the event of a fall or any similar health related emergencies. Delay in care from a health-related emergency, such as a fall, can have both short-term and long-term health related impacts.

2.  Several Indoor Localization systems are context-based and are functional only in the specific environments in which they were developed. For instance, [26] was developed for factory environments, [27] was developed for indoor parking, [28] was developed for hospital settings, [29] was developed for tracking forklift trucks in industry-based settings, and [30] was developed for performing Indoor Localization in academic environments for taking attendance of students. The future of interconnected Smart Cities would consist of a host of indoor environments in the living and functional spaces of humans, which would be far more diverse, different, and complicated as compared to the environments described in [26–30]. The challenge is thus to develop a means for Indoor Localization that is not environment dependent and can be seamlessly deployed in any IoT-based setting irrespective of the associated context parameters and their attributes.

3.  In view of the average dimensions of the living spaces in Smart Homes, the RMSE of the existing Indoor Localization systems are still high and greater precision and accuracy for detection of indoor location in the need of the hour.

4.  A range of machine learning-based approaches—Random Forest, Artificial Neural Network, Decision Tree, Support Vector Machine, k-NN, Gradient Boosted Trees, Deep Learning, and Linear Regression, have been used by several researchers [9–25] for development of various types of Indoor Localization systems for IoT-based environments. Identification of the optimal machine learning model that can be used to develop the future of Indoor Localization systems, Indoor Positioning Systems, and Location-Based Services is highly necessary.

Addressing these above-mentioned challenges by exploring the intersections of Big Data, Machine Learning, Indoor Localization, Ambient Assisted Living, Internet of Things, Activity Centric Computing, Human–Computer Interaction, Pattern Recognition, and Assisted Living Technologies to contribute towards AAL in Smart Homes serves as the main motivation for this work.

## 3. Technology Review

This section briefly reviews RapidMiner, formerly known as Yet Another Learning Environment (YALE) [60], which we have used for the work presented in this paper. RapidMiner is a software tool that allows development and implementation of a wide range of Machine Learning, Data Science, Artificial Intelligence, and Big Data related algorithms and models. The initial version of this tool was developed back in 2001 at the Technical University of Dortmund. From 2006, a company called Rapid-I started implementing additional functionalities and features in the tool. A year later, the name of the software was changed from YALE to RapidMiner and six years from then, the name of the company was changed from Rapid-I to RapidMiner. As of current day, RapidMiner is used both for educational research and for development of commercial applications and products.

RapidMiner is available as an integrated development environment that consists of—(1) RapidMiner Studio, (2) RapidMiner Auto Model, (3) RapidMiner Turbo Prep, (4) RapidMiner Go, (5) RapidMiner Server, and (6) RapidMiner Radoop. For all the work related to the methodologies proposed in this paper, we used RapidMiner Studio. For the remainder of this paper, wherever we have mentioned "RapidMiner", we have referred to "RapidMiner Studio" and not any of the other development environments associated with this software tool.

RapidMiner is developed as an open core model that provides a rich Graphical User Interface (GUI) to allow users to develop different kinds of applications, generate work-flows, and implement various algorithms. These applications, workflows, or algorithms are known as "processes" and they consist of multiple "operators". In a RapidMiner "process", each of its "operators" are associated with a specific functionality which is required for working of the "process". RapidMiner provides a range of built-in "operators" that can be directly used with or without any modifications for development of a specific "process". There is also a certain category of "operators" that can be used to modify the characteristic features of other "operators". The tool also allows developers to create their own "operators" and these can be shared and made available to all other users of RapidMiner via the RapidMiner Marketplace.

For development of any RapidMiner "process", the associated "operators" are always connected either in a linear fashion or in a hierarchical manner as shown in Figures 1 and 2, respectively. In Figure 1, 'Learner', 'Model Applier', and 'Evaluator' refer to different "operators" in RapidMiner. The inputs to these "operators" are shown by arrows pointing towards these respective "operators" and the outputs of these "operators" are shown by arrows pointing away from these respective "operators". For instance, the input to the 'Evaluator' "operator" is Example-Set and the output produced by this "operator" is the Performance Vector. Here, only three "operators" have been shown for representation of the linear arrangement amongst "operators", however, in an actual RapidMiner "process", the number of "operators" can vary as well as the specific "operators" could be different from the three "operators" shown in Figure 1. Similarly, a typical RapidMiner "process" is shown in Figure 2 which shows hierarchical arrangement amongst Methods 1, 2, and 3, each of which are "operators" Here, the "operators" 'Learner', 'Model Applier', and 'Evaluator' are also connected in a hierarchical manner. In an actual RapidMiner "process" the number and types of "operators" connected hierarchically could be different as compared to the number and types of "operators" shown in Figure 2.

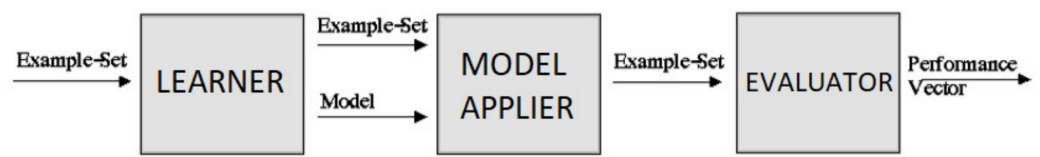

**Figure 1.** Layout of a typical linear "process" in RapidMiner Studio. This "process" shows linear arrangement amongst three RapidMiner "operators"—the 'Learner' operator, the 'Model Applier', and the 'Evaluator' [4].

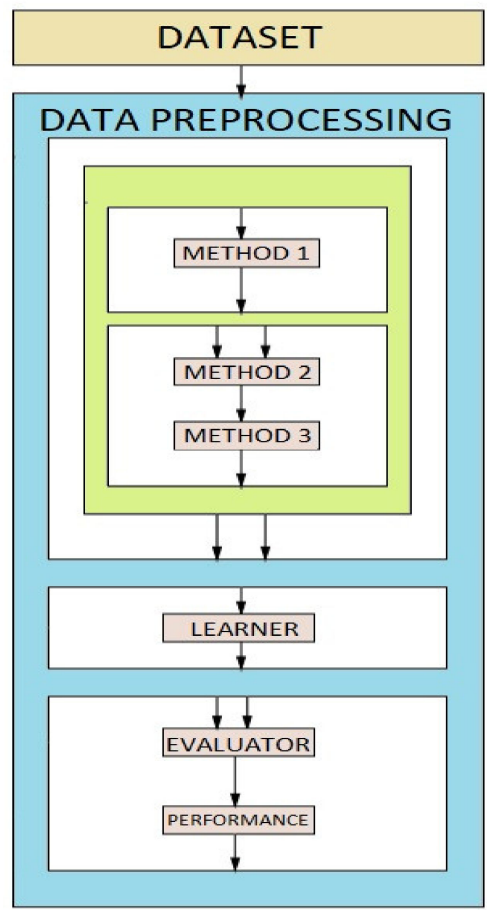

**Figure 2.** Layout of a typical hierarchical "process" in RapidMiner Studio. This "process" shows hierarchical arrangement amongst Methods 1, 2, and 3 each of which are "operators". The "operators"—'Learner', 'Model Applier', and 'Evaluator' are also connected in a hierarchical manner [4].

The following are some of the salient features of RapidMiner Studio:

1. It provides built-in "operators" with distinct functionalities that can be directly used or modified for development and implementation of Machine Learning, Data Science, Artificial Intelligence, and Big Data related algorithms and applications.
2. RapidMiner is developed using Java. This makes RapidMiner "processes" platform independent and Write Once Run Anywhere (WORA), which is a characteristic feature of Java.
3. The tool allows downloading multiple extensions for seamless communication and integration of RapidMiner "processes" with other software and hardware platforms.
4. Scripts written in any programming language, such as Python and R can also be integrated in a RapidMiner "process" to add additional functionalities to the same.
5. The tool allows development of new "operators" and seamless sharing of the same via the RapidMiner community.

6.  It also consists of "operators" that allow this software tool to connect with social media profiles of the user, such as Twitter and Facebook, to extract tweets, comments, posts, reactions, and related social media activity.

RapidMiner is developed as a client-server model. The server is made available as on-premise, in public or as a private cloud infrastructure. There are two different versions of RapidMiner available—the free version and the commercial version. The primary difference between these two versions is that the free version has a data processing limit of 10,000 rows for any "process". For all the work presented in this paper, the free version of RapidMiner 9.8.001 was downloaded and installed on a Microsoft Windows 10 Computer with Intel (R) Core (TM) i7-7600U CPU @ 2.80GHz, 2904 MHz, 2 Core(s) and 4 Logical Processor(s). The datasets that were studied and analyzed in this paper did not have more than 10,000 rows of data so this limitation of the free version of RapidMiner did not have any effect on the methodologies or the associated results and findings.

There are a few similar software tools that allow seamless development and implementation of Machine Learning, Data Science, Artificial Intelligence, and Big Data related algorithms and applications. Two such software tools which are very popular are—(1) Waikato Environment for Knowledge Analysis (WEKA) [61] and (2) MLC++ [62]. WEKA, developed in Java, allows development and implementation of various kinds of machine learning methods—such as regression, classification, feature selection, cross-validation, and bootstrapping. MLC++ is a C++ library that allows development and implementation of only supervised machine learning algorithms. The primary limitation of both WEKA and MLC++ is that they do not allow nesting of "operators" as supported by RapidMiner. In RapidMiner, "operator" nesting can be done either in a linear form or in a hierarchical form as shown in Figures 1 and 2, respectively. The only means for implementing such a feature in WEKA or MLC++ is by creating duplicate copies of the original dataset. However, this process is time consuming and requires a lot of memory space. To add to the above limitation of WEKA, it needs the current dataset to be available in the main memory of the system in which it is being executed. This also contributes towards consumption of computer memory. Neither WEKA nor MLC++ allow inclusion of programming scripts, such as scripts written in Python or R in their respective applications. To add to the above, MLC++ is not platform independent and neither does it have the WORA feature. In view of the above limitations of WEKA and MLC++ and the contrasting salient features of RapidMiner, RapidMiner was used for development of all the methodologies outlined in this paper.

## 4. Development of the Proposed Methodologies for Indoor Localization

For development of the proposed methodologies for Indoor Localization for AAL during ADLs performed in an IoT-based environment, such as a Smart Home, we posit the following:

(a)  The Received Signal Strength indicator (RSSI) data coming from BLE scanners and BLE beacons can be studied and analyzed to detect the changes in a user's instantaneous location during different activities, which are a result of the varying user interactions with dynamic context parameters.

(b)  The dynamic changes in the spatial configurations of a user during different activities can be interpreted by the analysis of the behavioral patterns that are localized and distinct for different activities.

(c)  Tracking and analyzing the user interactions with the context parameters along with the associated spatial information, by using a reference system, helps to detect the dynamic spatial configurations of the user.

Here, we use the concept of complex activity analysis for analysis of ADLs at a macro and micro level. A complex activity may be broadly defined as a collection of atomic activities (Ati), context attributes (Cti), core atomic activities ($\gamma$At), and core context attributes ($\rho$Ct) along with their characteristics [63]. The atomic activities refer to the small actions and tasks associated with an activity. The environment variables or parameters on which these tasks are performed are known as the context attributes. The core atomic activities refer to the atomic activities that are crucial for completion of the given complex activity and the context attributes on which these core atomic activities occur are known as the core context attributes. The atomic activities that are performed at the beginning and end of a given complex activity are known as start atomic activities (AtS) and end atomic activities (AtE), respectively. The associated context parameters are known as start context attributes (CtS) and end context attributes (CtE), respectively. This is further elaborated in Figure 3. Figure 4 describes a few atomic activities and complex activities in a typical environment [4]. Tables 1 and 2 show the complex activity analysis of two typical ADLs, Preparing Breakfast and Eating Lunch [4], studied in terms of the associated atomic activities, context attributes, core atomic activities, and core context attributes along with their associated weights, which can be determined by probabilistic reasoning principles [63]. As per [63], a greater weight of an Ati or Cti signifies greater relevance of the same towards the given complex activity. Therefore, weights of all $\gamma$At and $\rho$Ct are higher as compared to the rest of the Ati or Cti. The weights associated with the Ati, Cti, $\gamma$At, $\rho$Ct, AtS, AtE, CtS, and CtE are used to determine the threshold function of the given complex activity. The threshold function underlines the condition for completion of the complex activity [63].

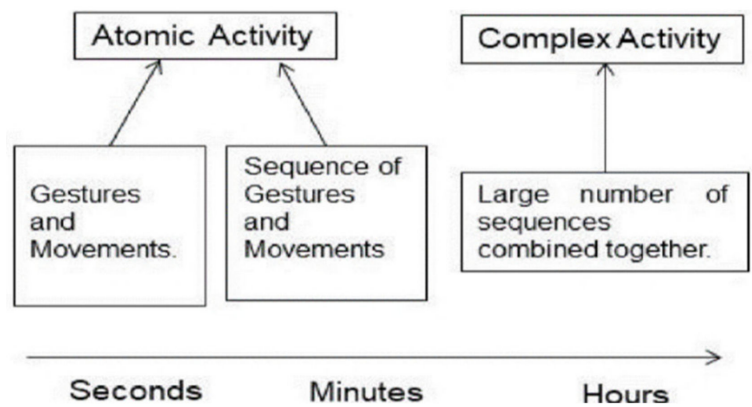

**Figure 3.** Representation of Atomic Activities and Complex Activities [4].

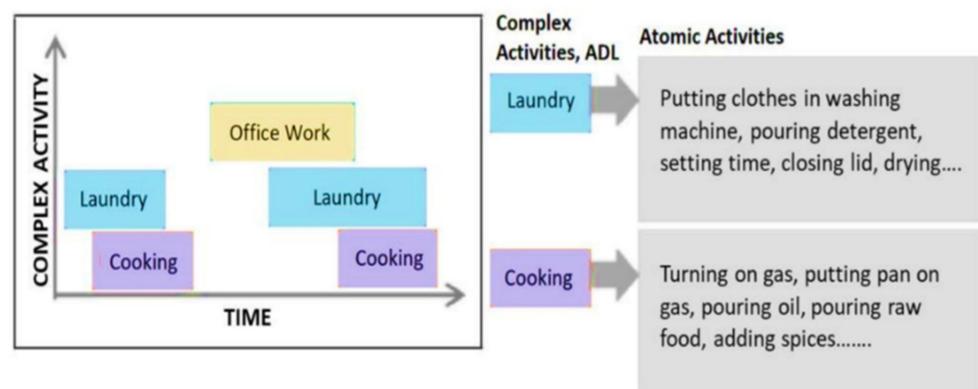

**Figure 4.** Description of Atomic Activities associated with two different Complex Activities in a typical environment [4].

**Table 1.** Analysis of a typical ADL, Preparing Breakfast (PB) from [4], studied in terms of the associated atomic activities, context attributes, core atomic activities, core context attributes and their threshold values.

| Complex Activity WCAtk (PB Atk)—PB (0.73) | |
| --- | --- |
| **Ati** | At1: Standing (0.10) <br> At2: Walking Towards Toaster (0.12) <br> At3: Putting bread into Toaster (0.15) <br> At4: Setting the Time (0.15) <br> At5: Turning off toaster (0.25) <br> At6: Taking out bread (0.18) <br> At7: Sitting Back (0.05) |
| **Cti** | Ct1: Lights on (0.10) <br> Ct2: Kitchen Area (0.12) <br> Ct3: Bread Present (0.15) <br> Ct4: Time settings working (0.15) <br> Ct5: Toaster Present (0.25) <br> Ct6: Bread cool (0.18) <br> Ct7: Sitting Area (0.05) |
| **AtS and CtS** | At1, At2, and Ct1, Ct2 |
| **AtE and CtE** | At6, At7, and Ct6, Ct7 |
| **$\gamma$At and $\rho$Ct** | At3, At4, At5, At6 and Ct3, Ct4, Ct5, Ct6 |

**Table 2.** Analysis of a typical ADL, Eating Lunch (EL) from [4], studied in terms of the associated atomic activities, context attributes, core atomic activities, core context attributes, and their threshold values.

| Complex Activity WCAtk (EL Atk)—EL (0.72) | |
| --- | --- |
| **Ati** | At1: Standing (0.08) <br> At2: Walking Towards Dining Table (0.20) <br> At3: Serving Food on a Plate (0.25) <br> At4: Washing Hand/Using Hand Sanitizer (0.20) <br> At5: Sitting Down (0.08) <br> At6: Starting to Eat (0.19) |
| **Cti** | Ct1: Lights on (0.08) <br> Ct2: Dining Area (0.20) <br> Ct3: Food Present (0.25) <br> Ct4: Plate Present (0.20) <br> Ct5: Sitting Options Available (0.08) <br> Ct6: Food Quality and Taste (0.19) |
| **AtS and CtS** | At1, At2, and Ct1, Ct2 |
| **AtE and CtE** | At5, At6, and Ct5, Ct6 |
| **$\gamma$At and $\rho$Ct** | At2, At3, At4 and Ct2, Ct3, Ct4 |

As can be seen from Table 1, for the complex activity of Preparing Breakfast in the environment described in [4], the atomic activities (Ati) are—Standing, Walking Towards Toaster, Putting bread into Toaster, Setting the Time, Turning off toaster, Taking out bread, and Sitting Back. The weights associated with these Ati are—0.10, 0.12, 0.15, 0.15, 0.25, 0.18, and 0.05, respectively. The associated context attributes (Cti) are—Lights on, Kitchen Area, Bread Present, Time settings working, Toaster Present, Bread cool, and Sitting Area. The weights associated with these Cti are—0.10, 0.12, 0.15, 0.15, 0.25, 0.18, and 0.05, respectively. These weights were assigned as per the probabilistic reasoning principles outlined in [63]. The Ati with the highest weights were identified as the Core Atomic Activities ($\gamma$At) as per the definition of $\gamma$At in [63]. The corresponding context attributes were identified as Core

Context Attributes (ρCt). Therefore, the list of γAt for this complex activity are—At3, At4, At5, and At6. Their corresponding context attributes—Ct3, Ct4, Ct5, and Ct6 were therefore considered as Core Context Attributes (ρCt). The Ati associated with the start and end of this complex activity, or in other words the Start Atomic Activities (AtS) and End Atomic Activities (AtE) are—At1, At2 and At6, At7, respectively. The corresponding context attributes—Ct1, Ct2 and Ct6, Ct7 were therefore considered as Start Context Attributes (CtS) and End Context Attributes (CtE), respectively. It is worth mentioning here that this complex activity analysis was performed based on the specific environment parameters described in [4] and this analysis can change if the same complex activity is performed in an environment which has a different set of environment parameters as compared to the environment described in [4].

As can be seen from Table 2, for the complex activity of Eating Lunch in the environment described in [4], the atomic activities (Ati) are—Standing, Walking Towards Dining Table, Serving Food on a Plate, Washing Hand/Using Hand Sanitizer, Sitting Down, and Starting to Eat. The weights associated with these Ati are—0.08, 0.20, 0.25, 0.20, 0.08, and 0.19, respectively. The associated context attributes (Cti) are—Lights on, Dining Area, Food Present, Plate Present, Sitting Options Available, and Food Quality and Taste. The weights associated with these Cti are—0.08, 0.20, 0.25, 0.20, 0.08, and 0.19, respectively. These weights were assigned as per the probabilistic reasoning principles outlined in [63]. The Ati with the highest weights were identified as the Core Atomic Activities (γAt) as per the definition of γAt in [63]. The corresponding context attributes were identified as Core Context Attributes (ρCt). Therefore, the list of γAt for this complex activity are—At2, At3, and At4. Their corresponding context attributes—Ct2, Ct3, and Ct4 were therefore considered as Core Context Attributes (ρCt). The Ati associated with the start and end of this complex activity or in other words the Start Atomic Activities (AtS) and End Atomic Activities (AtE) are—At1, At2 and At5, At6, respectively. The corresponding context attributes—Ct1, Ct2 and Ct5, Ct6 were therefore considered as Start Context Attributes (CtS) and End Context Attributes (CtE), respectively. It is worth mentioning here that this complex activity analysis was performed based on the specific environment parameters described in [4] and this analysis can change if the same complex activity is performed in an environment which has a different set of environment parameters as compared to the environment described in [4].

In the following sub sections, we outline the methodologies for development of the multimodal approaches for Indoor Localization for AAL for testing and evaluation of the above three hypotheses.

### 4.1. Indoor Localization from BLE Beacons and BLE Scanners Data during ADLs

The following are the steps for development of this proposed functionality:

i.    Set up an IoT-based environment, within a spatial context such as indoor layout of rooms with furniture's and appliances, using wearables and wireless sensors to collect the Big Data related to different ADLs. The associated representation scheme involves mapping the entire spatial location into non-overlapping 'activity-based zones', distinct to different complex activities, by performing complex activity analysis [63].

ii.   Analyze the ADLs in terms of the associated atomic activities, context attributes, core atomic activities, and core context attributes and their associated threshold values by probabilistic reasoning principles [63].

iii.  Infer the semantic relationships between the changing dynamics of atomic activities, context attributes, core atomic activities associated to different ADLs to study and interpret the spatial and temporal features of these ADLs.

iv.   Study the characteristics of the data coming from the wireless sensors to analyze the associated RSSI data from BLE beacons and BLE scanners, recorded during different ADLs, based on the user's proximity to the context attributes in each 'activity-based zone'. This information helps to infer the user's presence or absence in each of these 'activity-based zones'. For instance, when the user performs a typical complex activity—cooking using microwave, based on the user's proximity to the microwave, the user's presence can be deduced in a 'zone' where the microwave is present.

v.   Associate the relationships from (iii) with the characteristics of the RSSI data from BLE beacons and BLE scanners and map the entire IoT-based environment into non-overlapping 'activity-based zones', that are distinct to each ADL, by taking into consideration all possible complex activities that may be performed in the confines of the given IoT-based space. For instance, in a typical IoT-based environment [4], if the complex activities performed include—Watching TV, Using Laptop, Listening to Subwoofer, Using Washing Machine, Cooking Food, and Taking Shower; the associated 'activity-based zones' could be TV zone, Laptop zone, Subwoofer zone, Washing Machine zone, Cooking zone, and Bathroom zone. This inference of the respective 'zones' is based on the complex activity analysis [63] of all these activities as presented in [4].

vi.   Split the data into training set and test set and train a learning model to study these relationships and patterns in the data to detect the indoor location associated with the given ADL, based on detecting the user's presence or absence in a specific 'activity-based zone' at a specific point of time.

vii.   Analyze the performance characteristics of the learning model by using a confusion matrix.

Step (i) above involves setting up a Big Data collection methodology to study, track, and interpret the multimodal components of user interactions associated with different complex activities performed in each of these 'activity-based zones'. The data collection could be performed by using the Context-Driven Human Activity Recognition Framework that has already been developed, implemented, and tested at the Multimedia and Augmented Reality Lab, in the Department of Electrical Engineering and Computer Science at the University of Cincinnati. The results of the same were published in [64]. This framework was developed based on the work by Ma [65]. This framework, developed in the form of a software package, has multiple functionalities related to Big Data collection and we briefly review the same here.

First, it uses Microsoft Kinect Sensors to track the varying changes in the postures, gestures, and user behavior by performing skeletal tracking. In this process of skeletal tracking, the human body is represented in the form of a skeletal with 20 joint points and their associated characteristics. These joint points include—hip center, spine, shoulder center, head, left shoulder, left elbow, left wrist, left hand, right shoulder, right elbow, right wrist, right hand, left hip, left knee, left ankle, left foot, right hip, right knee, right ankle, and right foot. Characteristics of these joint points—such as joint point distance, joint point rotation, and joint point speed are tracked by this framework to detect and reason various movements and behavioral patterns for performing activity recognition. For instance, when a person is clapping, the distance between the joint point pairs (7,11) and (8,12) decreases and then increases in a periodic manner, where 7, 11, 8, and 12 represent the left wrist, right wrist, left hand, and right hand joint points, respectively. Based on these joint point characteristics, the framework can classify behaviors as Type-1 and Type-2. Behaviors associated with the lower limb are classified as Type-1 and behaviors associated with the upper limb are classified as Type-2, respectively. The Type-1 behaviors that can be recognized by this framework include—standing, walking, and sitting. The Type-2 behaviors that can be recognized by this framework include—waving, talking over cell phone, reading book or magazine, sleeping while seated, seated relaxed, and making hand gestures while talking. The third layer of the framework analyzes the relationships between changing behaviors associated with different activities by interpreting the user interactions with context parameters as well as the object behaviors in the user's spatial environment. It uses context-driven reasoning principles to perform complex activity recognition and analysis. To add, this layer can also track the sequence in which different complex activities are performed. The fourth layer of this framework allows capturing of social interactions and can perform all the functionalities of the above three layers while considering two users in the given IoT-based space at any point of time. This framework was developed as a Microsoft Windows-based application [64] that can seamlessly communicate and connect with both wireless and wearable sensors used to collect Big Data related to different complex activities. It consists of an intuitive user interface that shows the real-time Big Data coming from the IoT-based sensors as well as it shows the analysis of the same to deduce the associated activity being performed as per the methodologies outlined above.

For development of the proposed Big-Data driven methodology that studies the multimodal components of user interactions and analyzes the data from BLE beacons and BLE scanners to track a user's indoor location in a specific 'activity-based zone' during ADLs, we used an open-source dataset by Tabbakha et al. [66]. This dataset was chosen because its attributes were same as the real-time data that could be collected and analyzed by the Context-Driven Human Activity Recognition Framework [64] as outlined above. This dataset contains activity and human behavior related data collected from both wireless sensors and wearables in an IoT-based environment. The data attributes present in this dataset include the accelerometer data, gyroscope data, and the RSSI data obtained from BLE beacons and BLE scanners. The simulated smart home environment in which this data was collected consisted of four rooms or 'zones'—kitchen, bedroom, office, and toilet. For collecting the data as presented in this dataset [66], the authors developed a wearable device by using the Linkit 7697 and the MPU6050 sensors. This device was placed on the user's waist during each experimental trial. This wearable device tracked the behavior related information of the user as well as collected position related data with respect to the user's location in each of the four rooms or 'zones'. A BLE beacon was incorporated in this wearable and Raspberry Pi-based BLE scanners were installed at different locations of the IoT-based space. These scanners tracked the position of the user by sensing the BLE beacon and interpreting the associated RSSI data. Each of these rooms or 'zones' had one BLE scanner placed on or near a context attribute associated with the distinct complex activity that would be performed in that 'zone'. The scanners were placed on the kitchen table in the kitchen 'zone', on the bed in the bedroom 'zone', next to the working table in the 'office' zone, and next to the toilet door in the toilet 'zone'. The authors set the advertising interval of the BLE beacon to 100 ms with the transmitting power of up to $-30$ dBm. The scanner was programmed to report to the data collection server if the beacon was missing or in other words if the user was not present in that 'zone'. For such scenarios, the associated RSSI value was updated to $-120$ in the dataset to indicate that the BLE beacon was out of range. The behavior related data was collected by this wearable by tracking the acceleration data (along the X, Y, and Z axes) and gyroscope data (along the X, Y, and Z axes) associated with the user's movements. The data of the accelerometer and gyroscope from the wearable were sampled at the rate of 20 Hz. A total of 20 volunteers (10 males and 10 females) had participated in the experimental trials. The attributes that we used for development of this functionality included the RSSI data from BLE beacons and BLE scanners in the simulated environment and the location information that was associated with the different ADLs performed in this environment.

To study, analyze, and interpret these relationships that exist in the dataset, to detect the associated spatial context, i.e., to infer the location of the user within a specific 'activity-based zone' during ADLs, in the indoor room layout at a specific point of time, we used RapidMiner [60], because of its salient features and characteristics that make it highly suitable for development of such an application. These features and characteristics of RapidMiner as well as additional details about the relevance for selection of RapidMiner for this work were outlined in Section 3.

### 4.1.1. System Architecture of the Methodology for Indoor Localization from BLE Beacons and BLE Scanners Data during ADLs

The flowchart of this proposed methodology is shown in Figure 5. This figure outlines the operation of this methodology, as discussed in the previous section, at a broad level. Here, by "Split Data", we refer to splitting the data into training set and test set, with 80% data being selected for the training and the remaining 20% being selected for the testing. The "performance" in Figure 5 refers to evaluation of the performance of the k-NN approach when tested on the test dataset. This was evaluated by using a confusion matrix. Next, we outline how this flowchart was used for development of this methodology. We used an open-source dataset by Tabbakha et al. [66] for development of this methodology. This dataset was chosen because its attributes were same as the real-time data that could be collected and analyzed by the Context-Driven Human Activity Recognition Framework [64] as outlined above (Step i). The functionalities of this proposed approach, i.e., Steps (ii) to (vii), were developed and implemented in RapidMiner as a "process" as shown in Figure 6.

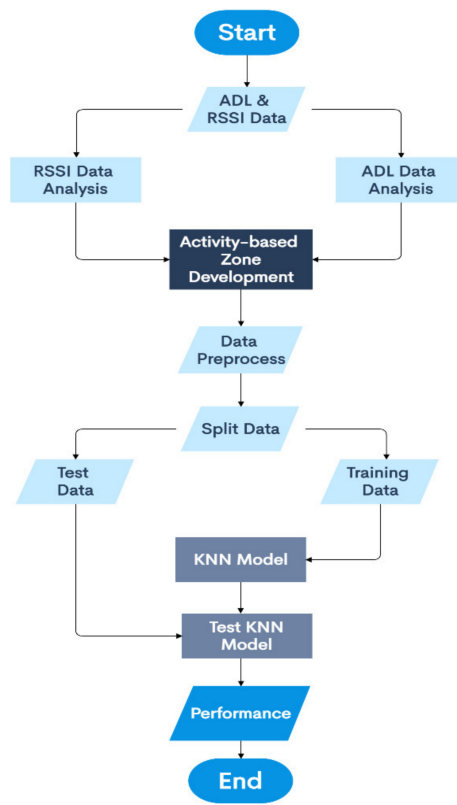

**Figure 5.** The flowchart for the proposed methodology for detection of indoor location in a specific 'activity-based zone' by analysis of the RSSI data coming from BLE beacons and BLE scanners during different ADLs.

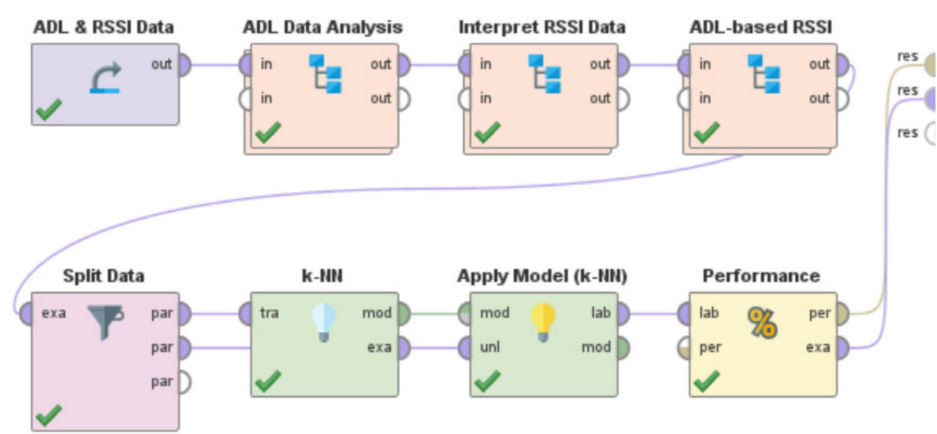

**Figure 6.** The RapidMiner "process" for detection of indoor location in a specific 'activity-based zone' by analysis of the RSSI data coming from BLE beacons and BLE scanners during different ADLs.

This "process" was developed as a combination of built-in "operators" and user defined "operators" in RapidMiner. An overview of built-in "operators" and user-defined "operators" in RapidMiner was presented in Section 3. We used the 'Dataset' "operator" to import this dataset into the RapidMiner "process". This "operator" was then renamed to 'ADL & RSSI Data' as for development of this "process" we needed to use only the activity-based data and the associated RSSI signals coming from BLE Beacons and BLE scanners during these ADLs. The semantic relationships between the changing dynamics of atomic activities, context attributes, core atomic activities, and core context attributes associated to different ADLs were then studied to interpret the spatial and temporal features of these ADLs (Step iii). This was done as per the methodology used to analyze complex activities (examples shown in Tables 1 and 2) and the associated "operator" that was developed was named as 'ADL Data Analysis'. The characteristics of the data coming from the sensors to analyze the associated RSSI data from BLE beacons and BLE scanners, recorded during different ADLs, were then studied and analyzed (Step iv) and the corresponding "operator" that was developed was named as 'Interpret RSSI Data'. Thereafter, we associated these relationships obtained from the 'ADL Data Analysis' "operator" with the characteristic features obtained from the 'Interpret RSSI Data' (Step v) to develop the 'activity-based zones'. We performed the same by developing an "operator"—'ADL-based RSSI'. Then, we used the built-in 'Split Data' "operator" to split the data into training set and test set with 80% of the data for training and 20% of the data for testing. A k-NN learning approach was used to develop the machine learning model which was tested on the test set by using the 'Apply Model (k-NN)' "operator". K-NN and 'Apply Model' are built-in "operators" in RapidMiner that can be directly used in any "process". Thereafter, we used the built-in 'Performance' "operator" in RapidMiner to evaluate the performance characteristics of the model in the form of a confusion matrix.

Figure 7 further clarifies the architecture of the proposed methodology as developed in RapidMiner [60]. This figure shows the flow of control depicting the sequence of operation of the different "operators" in this RapidMiner "process". As can be seen from Figure 7, the "operator" 'ADL & RSSI Data' is executed first, which is followed by the executions of the 'ADL Data Analysis', 'Interpret RSSI Data', 'ADL-based RSSI', 'Split Data', 'k-NN', 'Apply Model (k-NN)', and 'Performance' "operators", respectively. This RapidMiner "process", that studies the multimodal components of user interactions during ADLs and analyzes the data from BLE beacons and BLE scanners to track a user's indoor location in a specific 'activity-based zone'—which could be either the kitchen or the bedroom or the office or the toilet 'zone', achieved an overall performance accuracy of 81.36%. Further discussion of these results, the associated performance characteristics, and the rationale behind using confusion matrix for evaluation of this methodology are presented in Section 5.1.

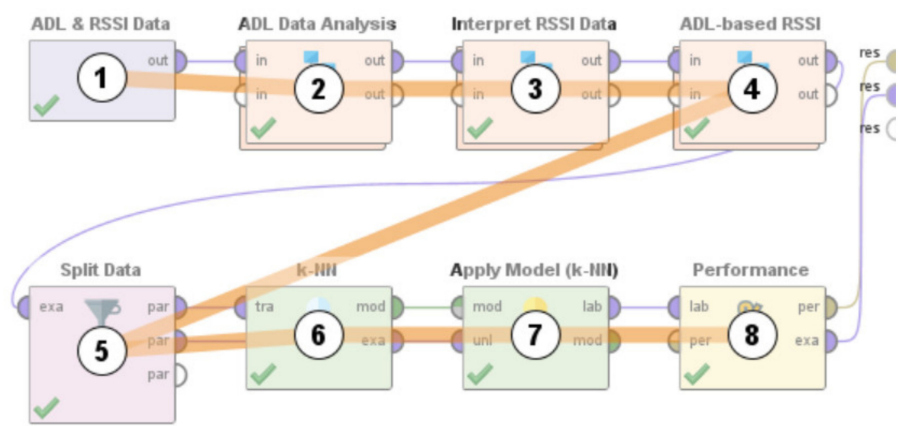

**Figure 7.** The flow of control showing the sequence of operation of the different "operators" in the RapidMiner "process" for detection of indoor location in a specific 'activity-based zone' by analysis of the RSSI data coming from BLE beacons and BLE scanners during different ADLs.

*4.2. Context Independent Indoor Localization from Accelerometer and Gyroscope Data*

The following are the steps for development of this functionality:

i.   Set up an IoT-based environment, within a spatial context such as indoor layout of rooms with furniture's and appliances, using wearables and wireless sensors to collect the Big Data related to different ADLs. The associated representation scheme involves mapping the entire spatial location into non-overlapping 'activity-based zones', distinct to different complex activities, by performing complex activity analysis [63] as outlined in Section 4.1.

ii.  Analyze the ADLs in terms of the associated atomic activities, context attributes, core atomic activities, and core context attributes, and their associated threshold values based on probabilistic reasoning principles [63].

iii. Infer the semantic relationships between the changing dynamics of atomic activities, context attributes, core atomic activities, and core context attributes along with the associated spatial and temporal information.

iv.  Study and analyze the semantic relationships between the accelerometer data (in X, Y, and Z directions), gyroscope data (in X, Y, and Z directions) and the associated atomic activities, context attributes, core atomic activities, and core context attributes within each 'activity-based zone'.

v.   Study and analyze the semantic relationships between the accelerometer data (in X, Y, and Z directions), gyroscope data (in X, Y, and Z directions) and the associated atomic activities, context attributes, core atomic activities, and core context attributes across different 'activity-based zones' based on the sequence in which the different ADLs took place and the related temporal information.

vi.  Integrate the findings from (iv) and (v) to interpret the interrelated and semantic relationships between the accelerometer data and the gyroscope data with respect to different ADLs performed in all the 'activity-based zones' in the given IoT-based space.

vii. Split the data into training set and test set and develop a machine learning-based model to detect the location of a user, in terms of these spatial 'zones' based on the associated accelerometer data (in X, Y, and Z directions) and gyroscope data (in X, Y, and Z directions).

viii. Evaluate the performance characteristics of the model by using a confusion matrix.

4.2.1. System Architecture of the Methodology for Context Independent Indoor Localization from Accelerometer and Gyroscope Data

The flowchart of the proposed methodology is shown in Figure 8. For convenient representation in this flowchart, accelerometer has been represented as "Acc" and gyroscope has been represented as "Gyro". This figure outlines the operation of this methodology, as discussed in the previous section, at a broad level. Here, by "Split Data", we refer to splitting the data into training set and test set, with 70% data being selected for the training and the remaining 30% being selected for the testing. The "performance" in Figure 8 refers to evaluation of the performance of the Random Forest approach when tested on the test dataset. This was performed by using the confusion matrix. Next, we outline the steps that we followed for implementation of this methodology as a RapidMiner process. As outlined in Section 4.1, for implementation of Step (i) above and for collection of Big Data related to ADLs, the Context-Driven Human Activity Recognition Framework could be used, that has already been developed, implemented, and tested at the Multimedia and Augmented Reality Lab, in the Department of Electrical Engineering and Computer Science at the University of Cincinnati. The results of the same were published in [64]. As this dataset by Tabbakha et al. [66] already had the data that we could have collected by setting up this data collection framework, so we used this dataset for development of the remaining functionalities of this methodology from Step (ii) in the form of a RapidMiner "process" as shown in Figure 9 by following the flowchart shown in Figure 8.

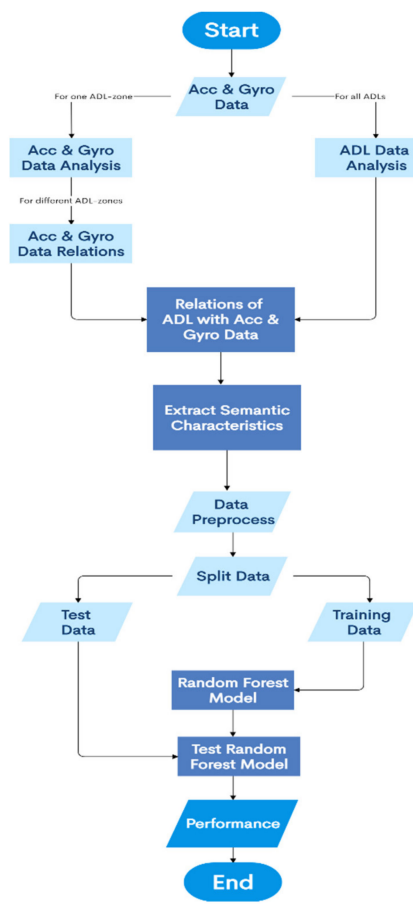

**Figure 8.** Flowchart of the proposed methodology for detection of indoor location based on the varying accelerometer and gyroscope data associated with distinct behavioral patterns related to distinct ADLs performed in distinct 'activity-based zones'.

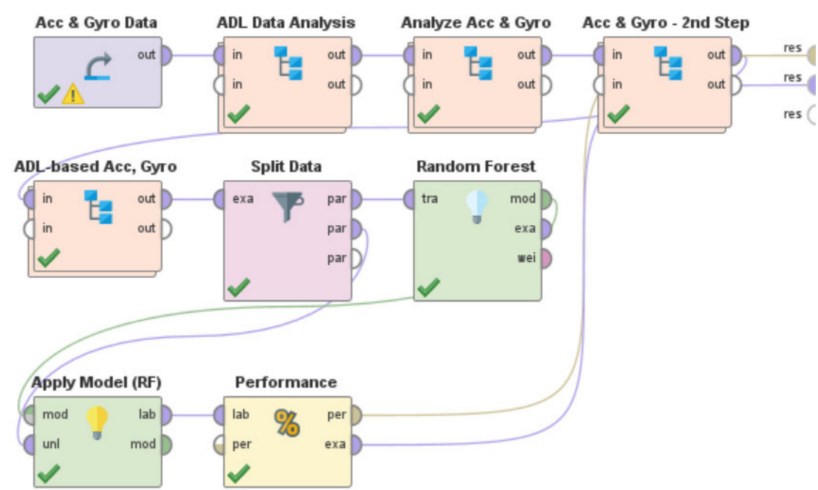

**Figure 9.** The RapidMiner "process" for detection of indoor location based on the varying accelerometer and gyroscope data associated with distinct behavioral patterns related to distinct ADLs performed in distinct 'activity-based zones'.

This "process" was developed as a combination of built-in "operators" and user defined "operators" in RapidMiner. An overview of built-in "operators" and user-defined "operators" in RapidMiner was presented in Section 3. We used the 'Dataset' "operator" to import this dataset into the RapidMiner "process". This "operator" was then renamed to 'Acc & Gyro Data' as for development of this "process" as we needed to use only the activity-based data and the associated accelerometer and gyroscope data. The semantic relationships between the changing dynamics of atomic activities, context attributes, core atomic activities, and core context attributes associated to different ADLs were then studied to interpret the spatial and temporal features of these ADLs (Step iii). This was done as per as the methodology used to analyze complex activities (examples shown in Tables 1 and 2) and the associated "operator" that was developed was named as 'ADL Data Analysis'. The functionality to study the semantic relationships between the accelerometer data (in X, Y, and Z directions), gyroscope data (in X, Y, and Z directions) and the associated atomic activities, context attributes, core atomic activities, and core context attributes within each 'activity-based zone' was then developed (Step iv) in the form of an "operator" which we named as 'Analyze Acc & Gyro'. Thereafter, we developed the functionality to study the semantic relationships between the accelerometer data (in X, Y, and Z directions), gyroscope data (in X, Y, and Z directions), and the associated atomic activities, context attributes, core atomic activities, and core context attributes across different 'activity-based zones' based on the sequence in which the ADLs took place as well as the associated temporal information, in the form of an "operator" which we named as 'Acc & Gyro 2nd Step' (Step v). The characteristic features of these "operators" were then merged to develop the functionality to interpret the interrelated and semantic relationships between the accelerometer data and the gyroscope data with respect to different ADLs performed in all the 'activity-based zones' (Step vi). This was done by developing an "operator" which we named as 'ADL-based Acc, Gyro'. Then, we used the built-in 'Split Data' "operator" to split the data into training set and test set with 70% of the data for training and 30% of the data for testing. A Random Forest-based learning approach was used to develop the machine learning model which was tested on the test set by using the 'Apply Model' "operator". 'Random Forest' and 'Apply Model' are built-in "operators" in RapidMiner that can be directly used in any "process". Thereafter, we used the built-in 'Performance' "operator" in RapidMiner to evaluate the performance characteristics of the model in the form of a confusion matrix.

Figure 10 further clarifies the architecture of the proposed methodology as developed in RapidMiner [60]. This figure shows the flow of control depicting the sequence of operation of the different "operators" in this RapidMiner "process". As can be seen from Figure 10, the "operator" 'Acc & Gyro Data' is executed first, which is followed by the executions of the 'ADL Data Analysis', 'Analyze Acc & Gyro', 'Analyze Acc & Gyro—2nd Step', 'ADL-based Acc, Gyro', 'Split Data', 'Random Forest', 'Apply Model (RF)', and 'Performance' "operators", respectively.

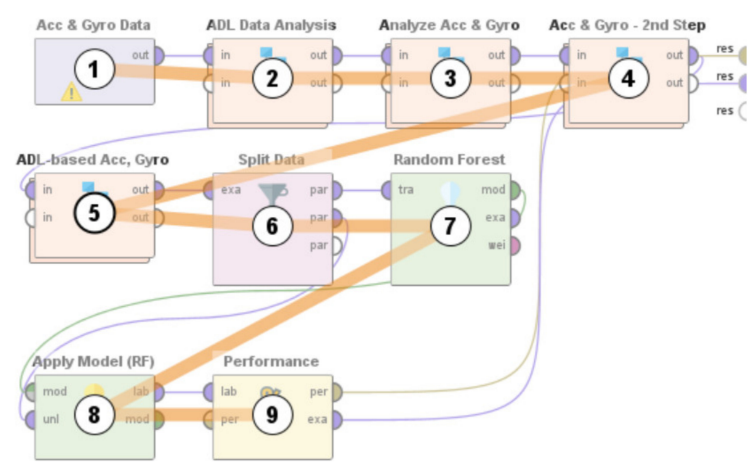

**Figure 10.** The flow of control showing the sequence of operation of the different "operators" in the RapidMiner "process" for detection of indoor location based on the varying accelerometer and gyroscope data associated with distinct behavioral patterns related to distinct ADLs performed in distinct 'activity-based zones'.

This methodology is based on interpretation of the accelerometer and gyroscope data from diverse behavioral patterns to detect the 'zone-based' indoor location of a user in any IoT-based environment. Here, the 'zone-based' mapping of a user's location refers to mapping the user in one of the multiple 'activity-based zones' that any given IoT-based environment can be classified into based on the specific activity being performed by the user. This classification of any given environment can be performed by the 'ADL Data Analysis' "operator" by using the complex activity recognition and analysis principles. The functionality of this "operator", as described above, is neither environment specific nor context parameter specific, thus its methodology can be applied for spatial mapping of any given IoT-based space. The accelerometer and gyroscope data that are analyzed and interpreted by this approach are a result or function of human behavior—that can be studied in the form of associated postures, gestures, movements, and motions, found in any IoT-based environment. Analysis of such behavior by the 'Analyze Acc & Gyro', 'Acc & Gyro 2nd Step', and 'ADL-based Acc, Gyro' is thus not dependent on a specific set of context parameters local to any specific IoT-based setting. All built-in data analysis related "operators" in RapidMiner are developed in a way so that they can be applied to any kind of data and they do not need any specific features in the environment to be present for their operation or function. The other "operators" that are a part of this "process"—'Split Data', 'Random Forest', 'Apply Model', and 'Performance' are built-in "operators" in RapidMiner and are therefore context independent. To summarize, all the operators that were used to develop this RapidMiner process, shown in Figure 9, are associated with distinct functionalities and characteristics that are not a function of any specific context-based or environment-based features local to any specific environment. In other words, these "operators" and thus the entire RapidMiner "process" as shown in Figure 9, would function for analysis and interpretation of any kind of user interaction data for performing the Indoor Localization of the user in that environment, based on the associated behavioral patterns distinct to different 'zones' local to that environment. This upholds the context independent nature of the entire RapidMiner "process" and thus the proposed methodology in Section 4.2. When tested on a dataset, this methodology, as shown in Figure 9, achieved a performance accuracy of 81.13%. Further discussion of these results, the associated performance characteristics, and the rationale behind using confusion matrix for evaluation of this methodology are presented in Section 5.2.

### 4.3. Detection of the Spatial Coordinates of the User in any 'Activity-Based Zone'

The following are the steps for development of this functionality in the proposed framework:

i.   Set up an IoT-based environment, within a spatial context such as indoor layout of rooms with furniture's and appliances, using wearables, and wireless sensors to collect the Big Data related to different ADLs.

ii.  The associated representation scheme involves setting up a context-based reference system in the given IoT-based environment. This system would track the instantaneous X and Y coordinates of the user's position information with respect to the origin of this reference system.

iii. Study each ADL performed in a specific 'activity-based zone' in terms of the multi-modal user interactions performed on the context parameters local to that 'zone'. This involves studying the atomic activities, context attributes, core atomic activities, core context attributes, start atomic activities, start context attributes, end atomic activities, and end context attributes.

iv.  For each of these user interactions with the context parameters, track the spatial configurations and changes in the user's position information, by using this reference system.

v.   Study the changes in the instantaneous spatial configurations of the user as per this reference system with respect to the dynamic temporal information associated with each user interaction performed in the given 'activity-based zone'.

vi.    Study and record all the user interactions as per (v), specific to the given ADL, in a given 'activity-based zone'.

vii.    Split the data into training set and test set and use the training set to train a machine learning-based model for detection of the varying X and Y coordinates of the user's position information in any 'activity-based' zone, as per the dynamic user interactions with context parameters.

viii.    Evaluate the performance characteristics of the model by using the root mean squared error method.

### 4.3.1. System Architecture of the Methodology for Detection of the Spatial Coordinates of the User in any 'Activity-Based Zone'

The flowchart for the proposed methodology is shown in Figure 11. This figure outlines the operation of this methodology, as discussed in the previous section, at a broad level. The distances from the three Bluetooth beacons that were used to develop the context-based reference system are represented as Distance A, Distance B, and Distance C, respectively. These distances were measured in meters. The actual X and Y coordinates of the user were measured in centimeters with an accuracy of $+/-1$ cm in the dataset [67] by analyzing the user's relative position with respect to these three beacons at a given point of time by using this reference system. In this figure, by "Split Data", we refer to splitting the data into training set and test set, with 70% data being selected for the training and the remaining 30% being selected for the testing. The "performance" in Figure 11 refers to evaluation of the performance of the machine learning approach when tested on the test dataset. This was evaluated by using the RMSE method, where RMSE errors were calculated separately in the X-direction and Y-direction as per ISO/IEC 18305:2016 [31]. This learning approach has been developed as a Random Forest model in this section. However, instead of a Random Forest model, other learning approaches such as Artificial Neural Network, Decision Tree, Support Vector Machine, k-NN, Gradient Boosted Trees, Deep Learning, and Linear Regression can also be seamlessly used for development of this methodology by following the flowchart shown in Figure 11. In Section 6, we have presented a comparative study where we implemented all these machine learning models—Random Forest, Artificial Neural Network, Decision Tree, Support Vector Machine, k-NN, Gradient Boosted Trees, Deep Learning, and Linear Regression for development of this methodology and compared their performance characteristics to deduce the optimal learning model for development of such an Indoor Localization system for detection of the spatial coordinates of the user in any 'activity-based zone'.

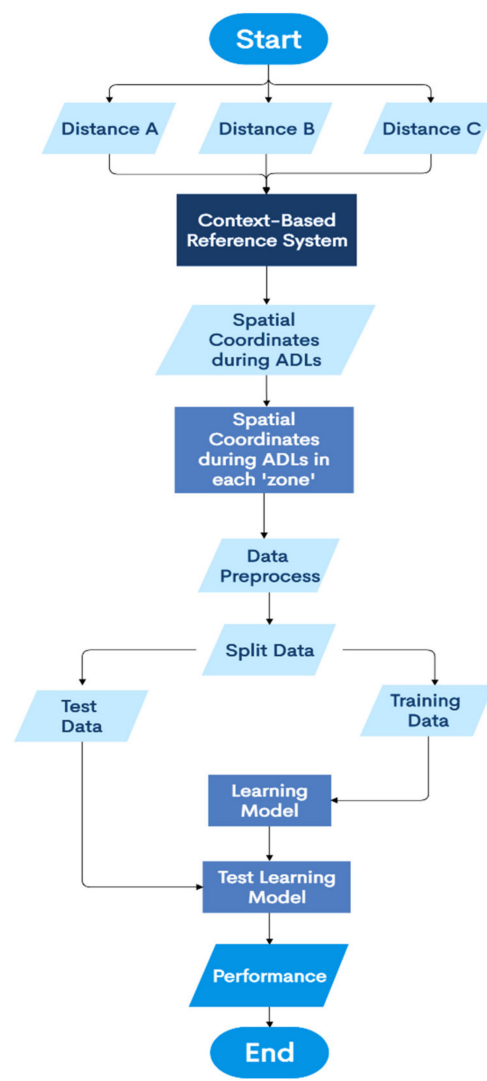

**Figure 11.** Flowchart for development of the methodology for detection of the varying X and Y coordinates of the user's position in any 'activity-based' zone, as per the dynamic user interactions with context parameters related to different activities.

Next, we outline the steps for development of the methodology as a RapidMiner process. As outlined in Section 4.1, for implementation of the Steps (i) and (ii) above and for collection of Big Data, the Context-Driven Human Activity Recognition Framework could be used, that has already been developed, implemented, and tested at the Multimedia and Augmented Reality Lab, in the Department of Electrical Engineering and Computer Science at the University of Cincinnati. The results associated with the same were published in [64]. As this dataset [67] already had the data and the corresponding data attributes, that we could have collected by setting up this data collection framework, so we used this dataset for development of the remaining functionalities of this methodology from Step (iii) in the form of a RapidMiner "process", as shown in Figure 12. This "process" was developed as a combination of built-in "operators" and user defined "operators" in RapidMiner. An overview of built-in "operators" and user-defined "operators" in RapidMiner was presented in Section 3.

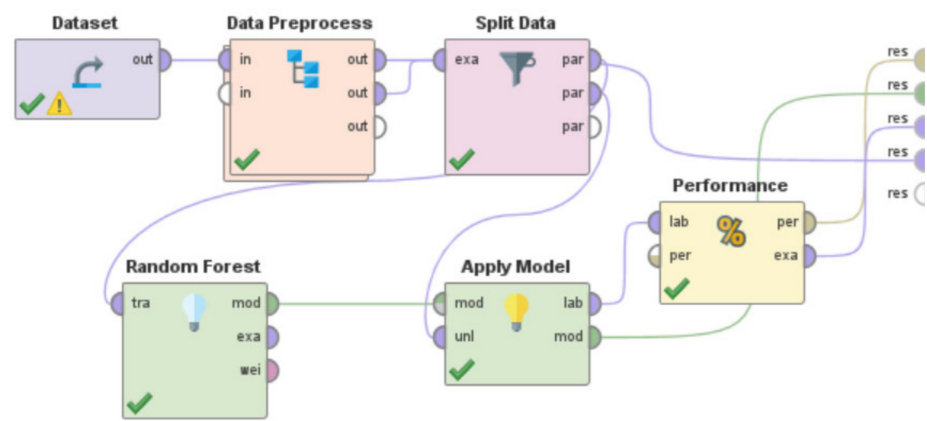

**Figure 12.** The RapidMiner "process" for detection of the varying X and Y coordinates of the user's position in any 'activity-based' zone, as per the dynamic user interactions with context parameters related to different activities.

This dataset used a reference system that was based out of comparing the user's spatial configuration, in terms of the actual distances—Distance A, Distance B, and Distance C, with respect to 3 Bluetooth beacons while analyzing the associated temporal information. This helped to detect the actual X and Y coordinates of the user in the given IoT-based space. The data consisted of 250 rows. We used RapidMiner [60] to develop a "process" to implement this functionality and evaluated its performance by using this dataset. The same version of RapidMiner and the same computer, as outlined in Section 3, were used for development of this "process". We used the 'Dataset' "operator" to import this dataset into the RapidMiner "process". The 'Data-Preprocess' "operator" was developed and then used to perform multiple preprocessing steps (Steps iii to vi) prior to splitting the data for training and testing. We used 70% of the data for training and the rest was used for testing. This data splitting was performed by the built in 'Split Data' "operator". A Random Forest learning approach was used to train the model by using the built-in 'Random Forest' "operator". The 'Apply Model' "operator", another built-in "operator", was used to apply the learning model on the test data and its performance characteristics were evaluated by using the 'Performance' "operator" in RapidMiner. We used the root mean square error (RMSE) method of evaluating the performance characteristics as per ISO/IEC18305:2016 [31]. This RapidMiner "process" is shown in Figure 12.

Figure 13 further clarifies the architecture of this proposed methodology as developed in RapidMiner [60]. This figure shows the flow of control depicting the sequence of operation of the different "operators" in this RapidMiner "process". As can be seen from Figure 13, the "operator" 'Dataset' is executed first, which is followed by the executions of the 'Data Preprocess', 'Split Data', 'Random Forest', 'Apply Model', and 'Performance' "operators", respectively. The RMSE for detection of X and Y coordinates of the user's position were found to be 5.85 cm and 5.36 cm, respectively. The Horizontal Error, as defined in ISO/IEC18305:2016 [31], was found to be 7.93 cm. Further discussion of these results, the associated performance characteristics, and the rationale behind using RMSE for evaluation of this methodology are presented in Section 5.3.

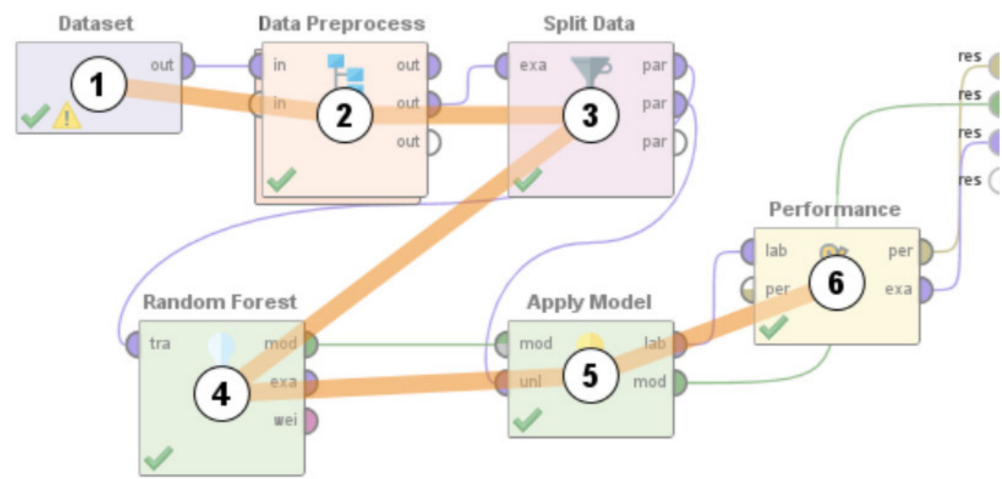

**Figure 13.** The flow of control showing the sequence of operation of the different "operators" in the RapidMiner "process" for detection of the varying *X* and *Y* coordinates of the user's position in any 'activity-based' zone, as per the dynamic user interactions with context parameters related to different activities.

## 5. Results and Findings

### 5.1. Indoor Localization from BLE Beacons and BLE Scanners Data during ADLs

In this section we present and discuss the results associated with the development of the proposed Big-Data driven methodology that studies the multimodal components of user interactions and analyzes the data from BLE beacons and BLE scanners to track a user's indoor location in a specific 'activity-based zone' during Activities of Daily Living, to test our first hypothesis—"*The RSSI data coming from BLE scanners and BLE beacons can be studied and analyzed to detect the changes in a user's instantaneous location during different activities, which are a result of the varying user interactions with dynamic context parameters*", as outlined in Section 4.1.

Upon development of the RapidMiner "process" as shown in Figure 6 by following the methods for development of this functionality (Section 4.1), we first studied the RSSI data from BLE beacons and BLE scanners associated with the varying atomic activities, context attributes, core atomic activities, and core context attributes associated with each of these ADLs performed in the different 'activity-based zones'. This is shown in Figures 14–16. In each of these figures, the *X*-axis represents the specific rooms or 'zones' modelled in the simulated Smart Home. The *Y*-axis represents the BLE scanner readings in different rooms or 'activity-based zones'. For instance, in Figure 14, the *X*-axis represents the different locations and the *Y*-axis represents the RSSI data recorded by the different BLE scanners present in the environment. From the dataset, it was observed that a BLE scanner provided an RSSI value of $-120$ when the BLE beacon was far away from the scanner or was out of its range. Therefore, for any value greater than $-120$, it could be concluded that the person was in that room or 'activity-based zone'. This is represented on the *Y*-axis. For instance, in Figure 14, the RSSI data is greater than $-120$ for the kitchen sensors but equal to $-120$ for sensors in all other rooms. This infers the presence of the person in the kitchen area.

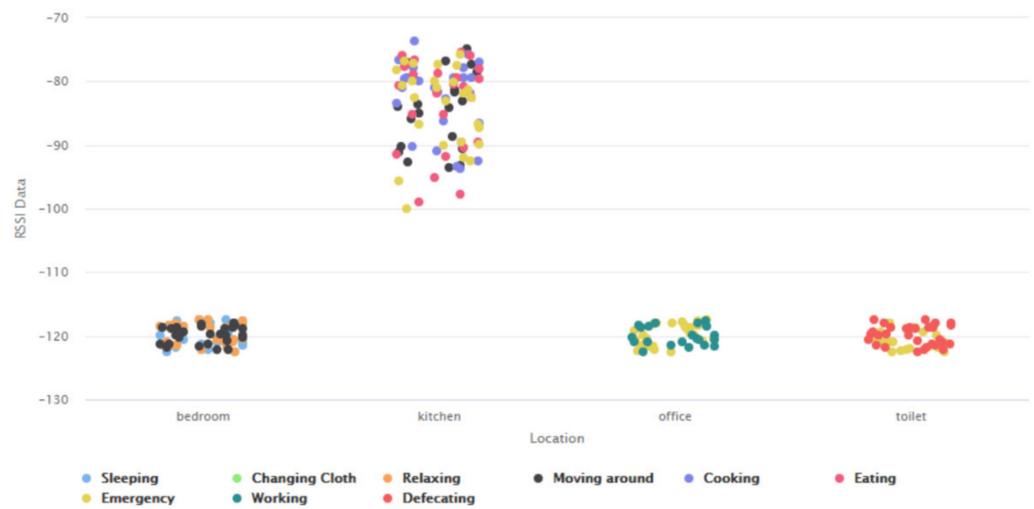

**Figure 14.** Analysis of RSSI data coming from different BLE scanners and BLE beacons when different activities were performed in the kitchen.

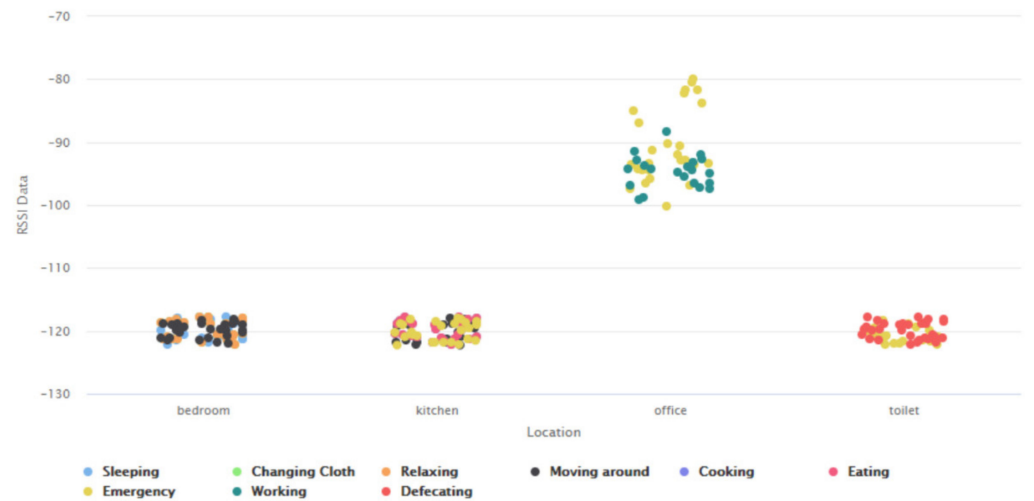

**Figure 15.** Analysis of RSSI data coming from different BLE scanners and BLE beacons when different activities were performed in the office area.

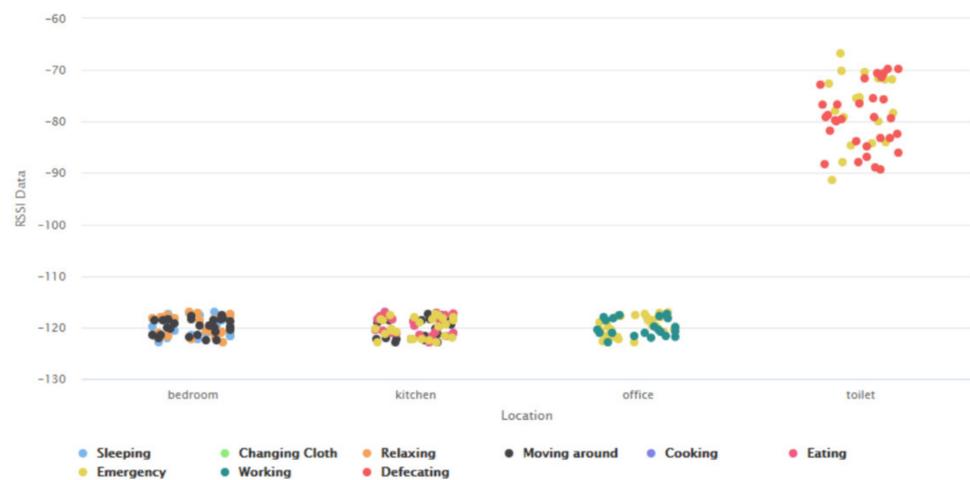

**Figure 16.** Analysis of RSSI data coming from different BLE scanners and BLE beacons when different activities were performed in the toilet.

The plots in each of these figures are color coded based on different complex activities that were performed by the user in each of these rooms or 'activity-based zones'—sleeping, changing clothes, relaxing, moving around, cooking, eating, working, defecating, and an emergency. An emergency constituted detecting the user in a lying position (either from a fall or unconsciousness) in an environment where a user is not supposed to lie down, for instance in the toilet. Figure 17 shows the output of the RapidMiner "process" (for the first 13 rows) which involved predicting the user's location in a specific 'activity-based zone' during different ADLs based on the associated RSSI data coming from the different BLE scanners and BLE beacons.

| Row No. | Location | prediction(Location) | confidence(bedroom) | confidence(kitchen) | confidence(office) | confidence(toilet) |
|---------|----------|---------------------|---------------------|---------------------|--------------------|--------------------|
| 1 | bedroom | bedroom | 0.800 | 0.200 | 0 | 0 |
| 2 | bedroom | kitchen | 0 | 1 | 0 | 0 |
| 3 | bedroom | bedroom | 0.607 | 0.393 | 0 | 0 |
| 4 | bedroom | bedroom | 0.801 | 0.199 | 0 | 0 |
| 5 | bedroom | bedroom | 0.804 | 0.196 | 0 | 0 |
| 6 | bedroom | bedroom | 0.614 | 0.386 | 0 | 0 |
| 7 | bedroom | kitchen | 0.195 | 0.805 | 0 | 0 |
| 8 | bedroom | kitchen | 0.423 | 0.577 | 0 | 0 |
| 9 | bedroom | bedroom | 0.605 | 0.395 | 0 | 0 |
| 10 | bedroom | bedroom | 0.613 | 0.387 | 0 | 0 |
| 11 | bedroom | bedroom | 0.814 | 0 | 0.186 | 0 |
| 12 | bedroom | bedroom | 0.802 | 0.198 | 0 | 0 |
| 13 | bedroom | bedroom | 0.824 | 0.176 | 0 | 0 |

**Figure 17.** Output of the RapidMiner "process" (first 13 rows) shown in Figure 6 to detect a user's location during different ADLs based on the RSSI data coming from BLE scanners and BLE beacons.

The output of the RapidMiner "process" assigned a confidence value for predicting the user's location in each of these 'activity-based zones'. The 'activity-based zone' with the highest confidence value was the final prediction of the model—in terms of predicting the user's location. For example, in Row number 3 of Figure 17, the confidence values of the model for the user's presence in the bedroom, kitchen, office and toilet 'zones' were, respectively, 0.607, 0.393, 0, and 0. As the confidence value of the model was highest for the bedroom 'zone', so, the final prediction of the model (third attribute from the left in Figure 17) was that the user was present in the bedroom. Table 3 consists of the description of all the attributes represented in the output shown in Figure 17.

**Table 3.** Description of the attributes of the output of the RapidMiner "process" shown in Figure 17.

| Attribute Name | Description |
|----------------|-------------|
| Row No | The row number in the output table |
| Location | The actual instantaneous zone-based location of the user |
| Prediction (Location) | The predicted instantaneous zone-based location of the user |
| Confidence (bedroom) | The degree of certainty that the user was in the bedroom |
| Confidence (kitchen) | The degree of certainty that the user was in the kitchen |
| Confidence (office) | The degree of certainty that the user was in the office area |
| Confidence (toilet) | The degree of certainty that the user was in the toilet |

The performance characteristics of this model, in terms of predicting whether the user was in the bedroom or kitchen or office area or toilet were evaluated by using a confusion matrix. Here, the data attribute being predicted by this methodology was the 'activity-based zone' and the associated values of the same were bedroom, kitchen, office, and toilet, as per the characteristics of the dataset used [64] and the associated functionalities of this approach (Section 4.1). As can be seen from Figure 17 and Table 3, none of these values of the 'activity-based zone' were numerical values. Even though ISO/IEC18305:2016 [31] recommends evaluating Indoor Localization systems either by using RMSE, or Mean of Error, or Covariance Matrix of Error, or Mean of Absolute Error, or Mean, or Standard Deviation of Vertical Error, etc.; such performance metrics can only be used when the predicted attribute is of numerical type—such as the numerical value of the X-coordinate, the numerical value of the Y-coordinate, the distance of the user from a specific reference point, etc. For non-numeric data types such performance metrics do not work. This is because one of the steps towards using the RMSE method of performance evaluation involves calculation of the arithmetic mean of the squares of a set of numbers [68] and similar mathematical operations are performed on the data when the other performance metrics as stated in ISO/IEC18305:2016 [31] are used. For non-numerical data neither can an arithmetic mean be computed nor can any mathematical operation be performed on the data. Evaluating the performance characteristics of an approach that involves prediction of non-numeric data by using a confusion matrix is a well-known practice in the field of machine learning, pattern recognition, data science, and their interrelated fields [69]. Therefore, we used a confusion matrix to study the performance characteristics of this methodology as proposed in Section 4.1.

The tabular representation and plot view of the performance characteristics, as obtained from RapidMiner, are shown in Figures 18 and 19, respectively. As can be observed from Figures 18 and 19, the model achieved an overall performance accuracy of 81.36%. The respective class recall values were 85.00%, 70.00%, 88.89%, and 90.00% for predicting the location of a user in bedroom, kitchen, office, and toilet, respectively. Further discussion about how this approach and the associated results and findings address multiple research challenges in this field is presented in Section 7.

**accuracy: 81.36%**

|  | true bedroom | true kitchen | true office | true toilet | class precision |
|---|---|---|---|---|---|
| pred. bedroom | 17 | 1 | 0 | 0 | 94.44% |
| pred. kitchen | 3 | 14 | 1 | 0 | 77.78% |
| pred. office | 0 | 0 | 8 | 1 | 88.89% |
| pred. toilet | 0 | 5 | 0 | 9 | 64.29% |
| class recall | 85.00% | 70.00% | 88.89% | 90.00% | |

**Figure 18.** A confusion matrix (tabular view) representing the performance accuracy of the Rapid-Miner "process" shown in Figure 6 to detect a user's location during different ADLs based on the RSSI data coming from BLE scanners and BLE beacons.

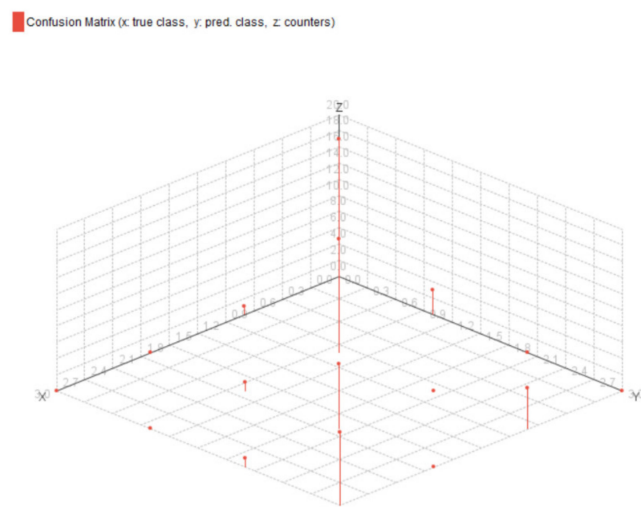

**Figure 19.** A confusion matrix (plot view) representing the performance accuracy of the RapidMiner "process" shown in Figure 6 to detect a user's location during different ADLs based on the RSSI data coming from BLE scanners and BLE beacons.

*5.2. Context Independent Indoor Localization from Accelerometer and Gyroscope Data*

In this section we present and discuss the results associated with the development of the proposed context independent approach that can interpret the accelerometer and gyroscope data from diverse behavioral patterns to detect the 'zone-based' indoor location of a user in any IoT-based environment, to test our second hypothesis—"*The dynamic changes in the spatial configurations of a user during different activities can be interpreted by the analysis of the behavioral patterns that are localized and distinct for different activities*", as outlined in Section 4.2. By using the RapidMiner "process" as shown in Figure 9 and by following the proposed functionalities of our framework (Section 4.2), we studied the variations of the accelerometer data (in X, Y, and Z directions) and gyroscope data (in X, Y, and Z directions) as per the variations in behavioral and user interaction patterns in the distinct 'activity-based zones' in the confines of the given IoT-based space. These variations in behavioral patterns were a result of the user performing different ADLs, characterized by distinct user interactions with the varying context parameters, in each of these 'activity-based zones'. This study is represented in Figures 20–25.

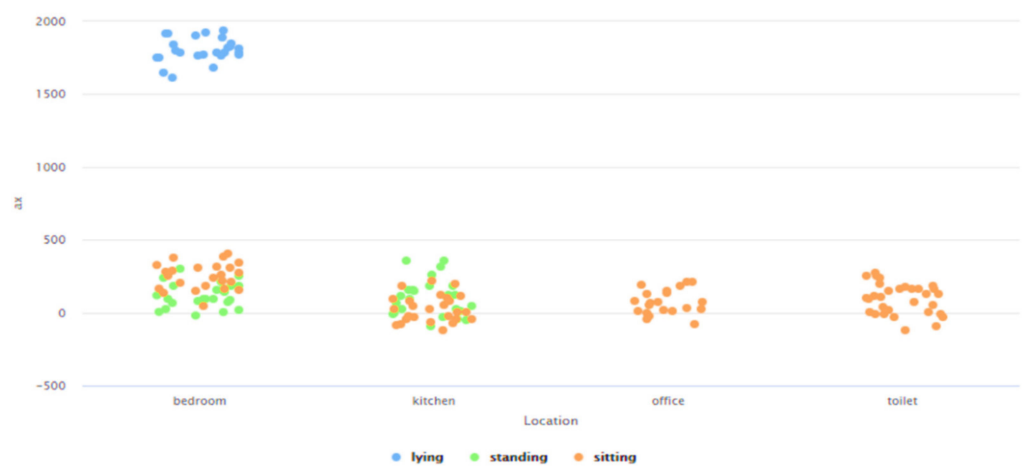

**Figure 20.** Analysis of the accelerometer data (in the X-direction) for different behavioral patterns associated with different ADLs performed in the given simulated smart home environment.

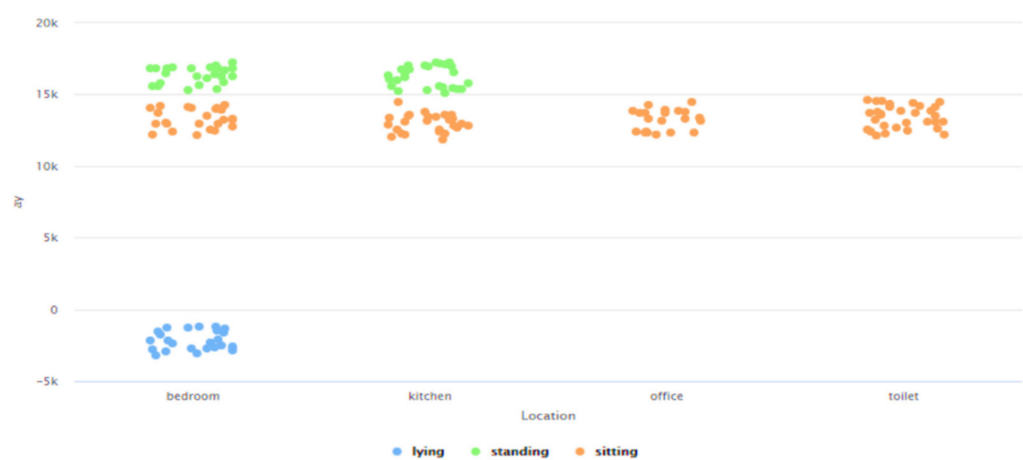

**Figure 21.** Analysis of the accelerometer data (in the Y-direction) for different behavioral patterns associated with different ADLs performed in the given simulated smart home environment.

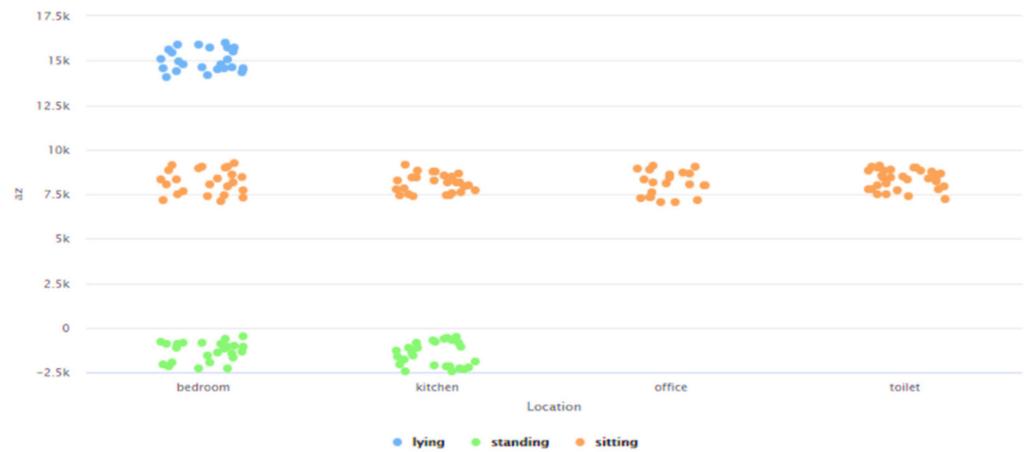

**Figure 22.** Analysis of the accelerometer data (in the Z-direction) for different behavioral patterns associated with different ADLs performed in the given simulated smart home environment.

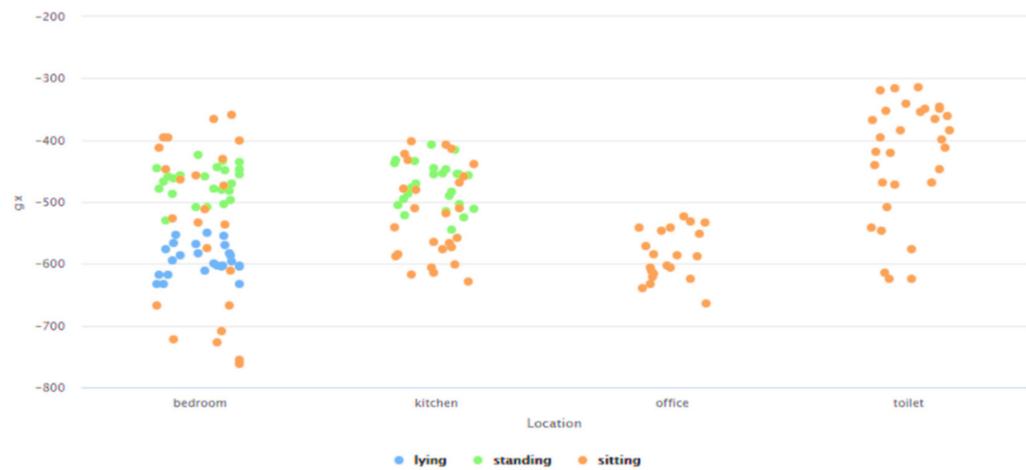

**Figure 23.** Analysis of the gyroscope data (in the X-direction) for different behavioral patterns associated with different ADLs performed in the given simulated smart home environment.

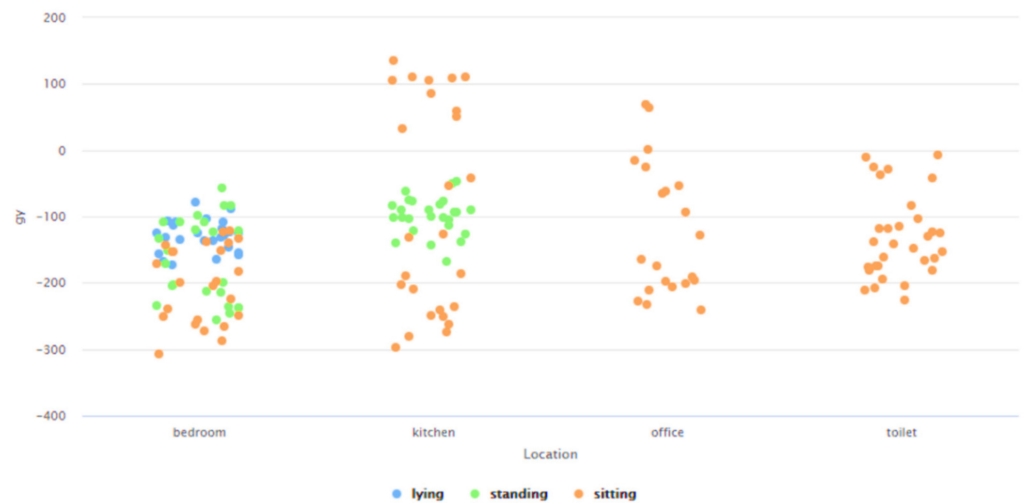

**Figure 24.** Analysis of the gyroscope data (in the Y-direction) for different behavioral patterns associated with different ADLs performed in the given simulated smart home environment.

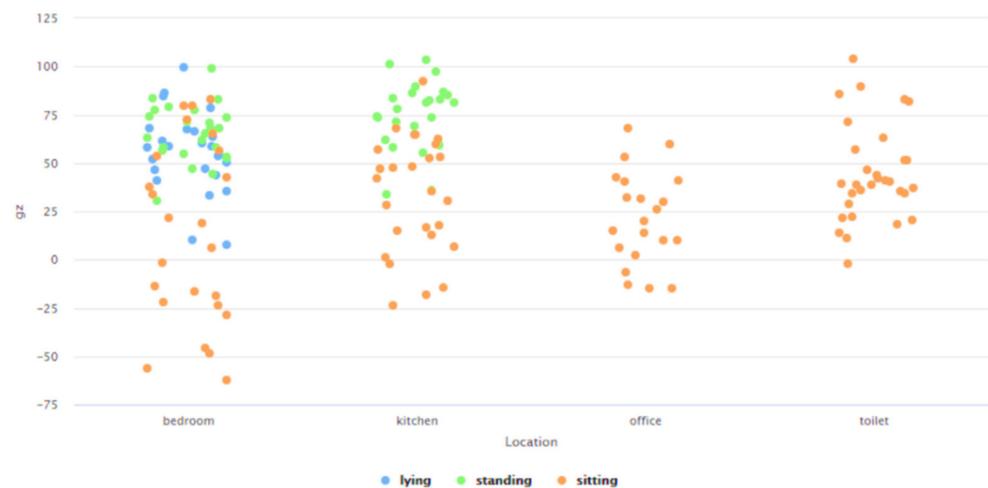

**Figure 25.** Analysis of the gyroscope data (in the Z-direction) for different behavioral patterns associated with different ADLs performed in the given smart home environment.

Figure 26 shows the output of the RapidMiner "process" (for the first 13 rows) which involved predicting the user's location in a specific 'activity-based zone' based on the associated accelerometer and gyroscope data. The output of the RapidMiner "process" assigned a confidence value for predicting the user's location in each of these 'activity-based zones'. The 'activity-based zone' with the highest confidence value was the final prediction of the model—in terms of predicting the user's location. For example, in Row number 7 of Figure 26, the confidence values of the model for the user's presence in the bedroom, kitchen, office, and toilet were, respectively, 0.985, 0.003, 0.010, and 0.002. As the confidence value of the model was highest for the bedroom so the final prediction of the model (third attribute from the left in Figure 26) was that the user was present in the bedroom. Table 4 consists of the description of all the attributes represented in the output shown in Figure 26.

| Row No. ↑ | Location | prediction(Location) | confidence(bedroom) | confidence(kitchen) | confidence(office) | confidence(toilet) |
|---|---|---|---|---|---|---|
| 1 | bedroom | bedroom | 0.988 | 0 | 0 | 0.012 |
| 2 | bedroom | bedroom | 0.995 | 0.001 | 0 | 0.004 |
| 3 | bedroom | bedroom | 0.987 | 0.001 | 0 | 0.013 |
| 4 | bedroom | bedroom | 0.995 | 0.001 | 0 | 0.004 |
| 5 | bedroom | bedroom | 0.995 | 0.001 | 0 | 0.004 |
| 6 | bedroom | bedroom | 0.995 | 0.001 | 0 | 0.004 |
| 7 | bedroom | bedroom | 0.985 | 0.003 | 0.010 | 0.002 |
| 8 | bedroom | bedroom | 0.971 | 0.013 | 0 | 0.016 |
| 9 | bedroom | bedroom | 0.993 | 0.004 | 0 | 0.004 |
| 10 | bedroom | bedroom | 0.945 | 0.043 | 0.010 | 0.002 |
| 11 | bedroom | bedroom | 0.834 | 0.139 | 0.013 | 0.014 |
| 12 | bedroom | bedroom | 0.777 | 0.197 | 0.013 | 0.014 |
| 13 | bedroom | bedroom | 0.662 | 0.297 | 0.018 | 0.023 |

**Figure 26.** Output of the RapidMiner "process" (first 13 rows) shown in Figure 9 to detect a user's location during different ADLs based on the associated accelerometer and gyroscope data.

**Table 4.** Description of the attributes of the output of the RapidMiner "process" shown in Figure 26.

| Attribute Name | Description |
|---|---|
| Row No | The row number in the output table |
| Location | The actual instantaneous zone-based location of the user |
| Prediction (Location) | The predicted instantaneous zone-based location of the user |
| Confidence (bedroom) | The degree of certainty that the user was in the bedroom |
| Confidence (kitchen) | The degree of certainty that the user was in the kitchen |
| Confidence (office) | The degree of certainty that the user was in the office area |
| Confidence (toilet) | The degree of certainty that the user was in the toilet |

The performance characteristics of this model in terms of predicting whether the user was in the bedroom or kitchen or office area or toilet were evaluated by using a confusion matrix.

Here, the data attribute being predicted by this methodology was the 'activity-based zone' and the associated values of the same were bedroom, kitchen, office, and toilet, as per the characteristics of the dataset used [64] and the associated functionalities of this approach (Section 4.2). As can be seen from Figure 26 and Table 4, none of these values of the 'activity-based zone' were numerical values. Even though ISO/IEC18305:2016 [31] recommends evaluating Indoor Localization systems either by using RMSE, or Mean of Error, or Covariance Matrix of Error, or Mean of Absolute Error, or Mean, or Standard Deviation of Vertical Error, etc.; such performance metrics can only be used when the predicted attribute is of numerical type—such as the numerical value of the X-coordinate, the numerical value of the Y-coordinate, the distance of the user from a specific reference point, etc. For non-numeric data types such performance metrics do not work. This is because one of the steps towards using the RMSE method of performance evaluation involves calculation of the arithmetic mean of the squares of a set of numbers [68] and similar mathematical operations are performed on the data when the other performance metrics as stated in ISO/IEC18305:2016 [31] are used. For non-numerical data neither can an arithmetic mean be computed nor can any mathematical operation be performed on the data. Evaluating the performance characteristics of an approach that involves prediction of non-numeric data by using a confusion matrix is a well-known practice in the field of machine learning, pattern recognition, data science, and their interrelated fields [69]. Therefore, we used a confusion matrix to study the performance characteristics of this methodology as proposed in Section 4.2.

The tabular representation and plot view of the performance characteristics as obtained from RapidMiner are shown in Figures 27 and 28, respectively. As can be observed from Figures 27 and 28, the model achieved an overall performance accuracy of 81.13%. The class recall values were 86.36%, 68.75%, 83.33%, and 88.89% for predicting the location of a user in bedroom, kitchen, office, and toilet, respectively. Further discussion about how this approach and the associated results and findings address multiple research challenges in this field is presented in Section 7.

**accuracy: 81.13%**

|                  | true bedroom | true kitchen | true office | true toilet | class precision |
|------------------|--------------|--------------|-------------|-------------|-----------------|
| pred. bedroom    | 19           | 4            | 0           | 0           | 82.61%          |
| pred. kitchen    | 1            | 11           | 0           | 1           | 84.62%          |
| pred. office     | 0            | 0            | 5           | 0           | 100.00%         |
| pred. toilet     | 2            | 1            | 1           | 8           | 66.67%          |
| class recall     | 86.36%       | 68.75%       | 83.33%      | 88.89%      |                 |

**Figure 27.** A confusion matrix (tabular view) representing the performance accuracy of the Rapid-Miner "process" shown in Figure 9 to detect a user's indoor location based on the associated accelerometer and gyroscope data.

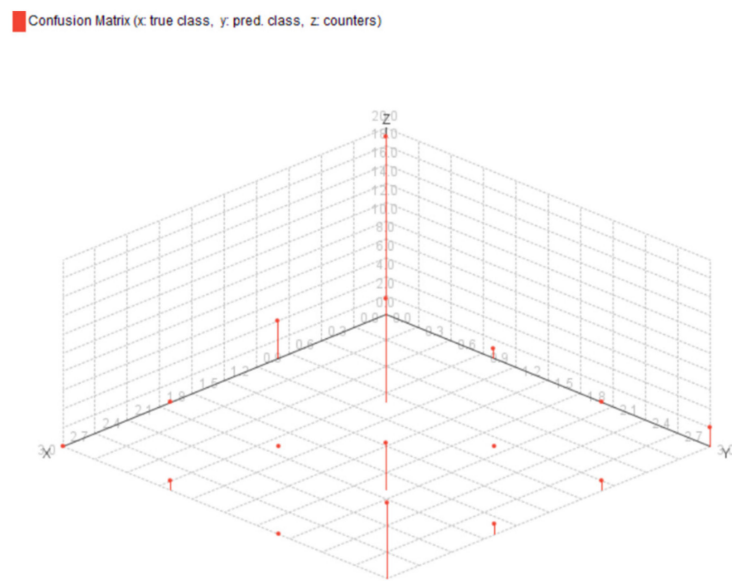

**Figure 28.** A confusion matrix (plot view) representing the performance accuracy of the RapidMiner "process" shown in Figure 9 to detect a user's indoor location based on the associated accelerometer and gyroscope data.

### 5.3. Detection of the Spatial Coordinates of the User in Any 'Activity-Based Zone'

In this section we present and discuss the results associated with the development of the proposed methodology to detect the spatial coordinates of a user's indoor position based on the associated user interactions with the context parameters and the user-centered local spatial context, by using a reference system, to test our third hypothesis—"*Tracking and analyzing the user interactions with the context parameters along with the associated spatial information, by using a reference system, helps to detect the dynamic spatial configurations of the user*", as outlined in Section 4.3. The process involved setting up a context-based reference system to track the user's location in the confines of the given IoT-based environment during different ADLs and consisted of multiple steps. The first step was to study each ADL performed in a specific 'activity-based zone' in terms of the multimodal user interactions performed on the context parameters local to that 'activity-based zone'. This involved studying the atomic activities, context attributes, core atomic activities, core context attributes, start atomic activities, start context attributes, end atomic activities, and end context attributes associated with all the complex activities. The second step involved tracking the spatial configurations and changes in the user's position information, by using the reference system (Section 4.3), during all the varying interactions with the context parameters. The methodology then studied the changes in the instantaneous spatial configurations of the user as per this reference temporal system with respect to the dynamic temporal information associated with each user interaction performed in any given 'activity-based zone' to train a machine learning model. Upon development of the RapidMiner "process" as shown in Figure 12, we first studied the dynamic changes in the user's distance from the three beacons that were used to develop the reference system of this dataset [67]. This is shown in Figures 29–32. The distances from these three beacons, measured in meters, are represented as Distance A, Distance B, and Distance C, respectively, in these figures. These distances were measured in meters and the actual X and Y coordinates of the user were measured in centimeters with an accuracy of $+/-1$ cm in the dataset [67].

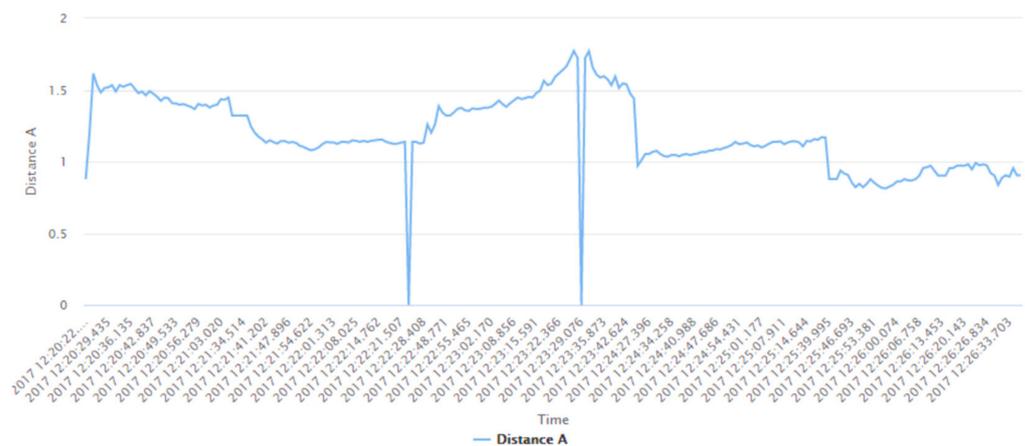

**Figure 29.** Analysis of the variation of Distance A (user's actual distance from one of the beacons) at different timestamps, based on the associated changes in behavioral patterns. Due to paucity of space the data associated with a few timestamps are shown here.

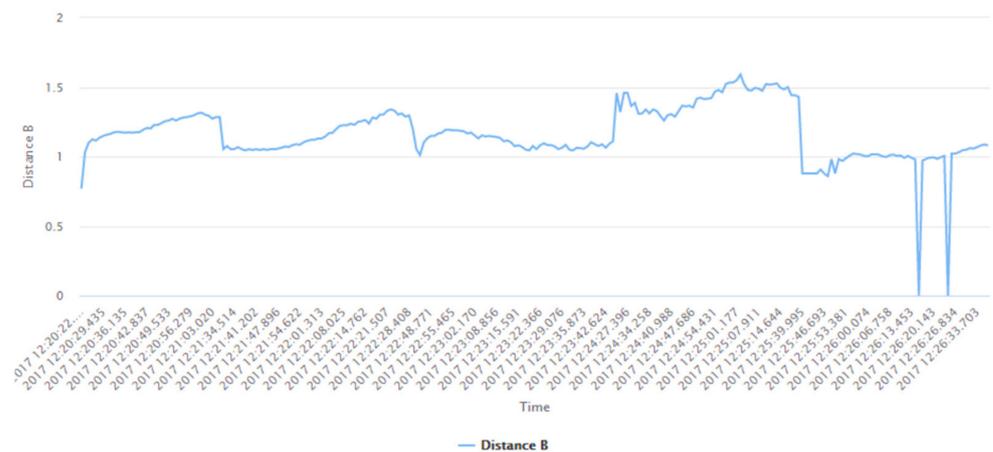

**Figure 30.** Analysis of the variation of Distance B (user's actual distance from one of the beacons) at different timestamps, based on the associated changes in behavioral patterns. Due to paucity of space the data associated with a few timestamps are shown here.

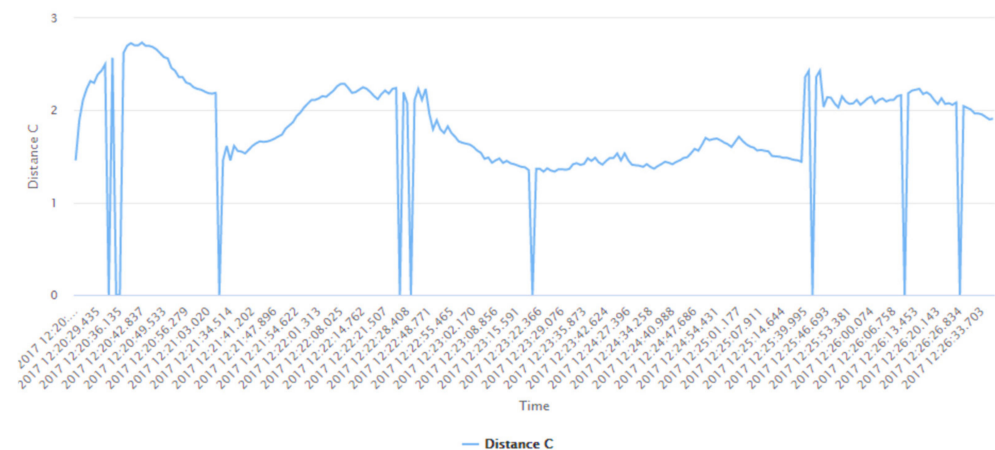

**Figure 31.** Analysis of the variation of Distance C (user's actual distance from one of the beacons) at different timestamps, based on the associated changes in behavioral patterns. Due to paucity of space the data associated with a few timestamps are shown here.

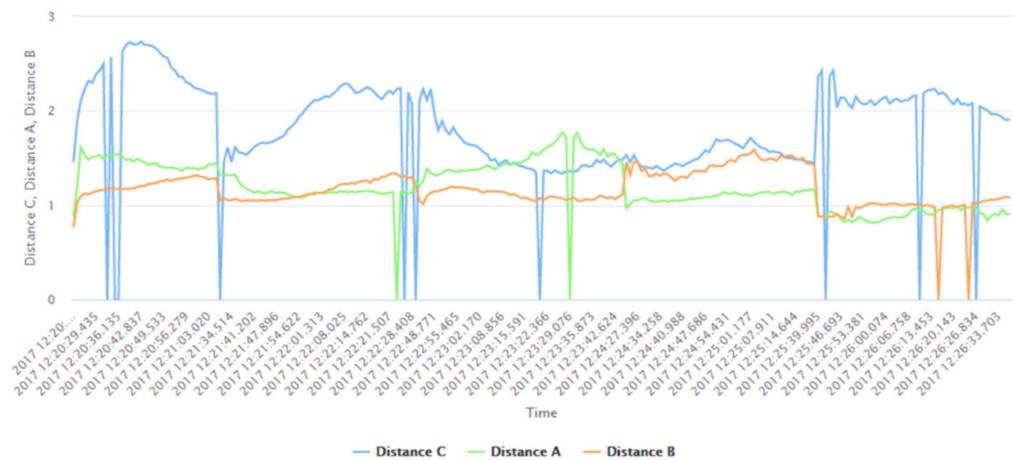

**Figure 32.** Analysis of the variation of Distances A, B, and C plotted together at different timestamps, based on the associated changes in behavioral patterns. Due to paucity of space the data associated with a few timestamps are shown here.

The Random Forest model that we developed in RapidMiner (Figure 12) consisted of 100 random trees, and used the least square criterion for splitting at each node. The maximum depth of a tree was 10. This learning model assigned weights to each of these distances—Distance A, Distance B, and Distance C to accurately track the user based on the provided reference system. The weights that the model associated with Distance A, Distance B, and Distance C were 0.531, 0.287, and 0.183, respectively, for determination of the X-coordinate of the user's location and these respective weights were 0.639, 0.170, and 0.191 for determination of the Y-coordinate of the user's location.

Figure 33 shows one of these random trees that was developed by the Random Forest model, as shown in Figure 12, and the reasoning-based description of this tree is shown in Figure 34. This random tree was associated with detecting different values of the X-coordinate of the user based on the associated rules at each node. We explain the working of the tree for one such detection here, when the user was located at the X-coordinate—79.000 as per this reference system. This is marked in blue in Figure 33. The comparison started at the topmost node—which for this tree was Distance A. This distance was lesser than or equal to 1.331 m, so the control moved to the right half of the tree. Then, it checked the value of Distance B, which was greater than 1.038 m, so it went to the left half of this node, where it checked the value of Distance A again. This value was greater than 1.078 m so it went to the right half and checked Distance B at the next node, which was less than or equal to 1.373 m, so the control moved to the right half of this node for checking Distance C. At this node, after performing the condition check, it moved to the left side of the node as Distance C was greater than 0.729 m. Then, the control compared the values of Distance A and Distance B with respect to a couple of more conditions at the respective child nodes to finally deduce the X-coordinate of the user as 79.000.

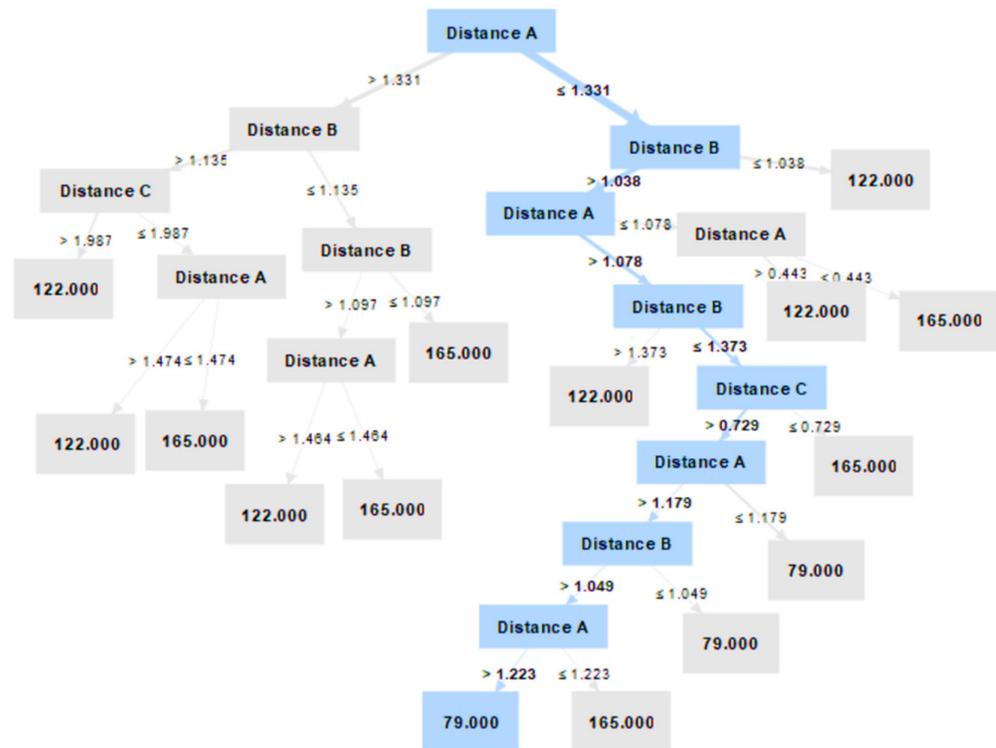

**Figure 33.** Representation of one of the Random Trees developed by the Random Forest-based RapidMiner "process" shown in Figure 12. This random tree was associated with detecting different values of the X-coordinate of the user based on the associated rules at each node.

```
Distance A > 1.344
|   Distance C > 2.147: 122.000 {count=30}
|   Distance C ≤ 2.147
|   |   Distance C > 0.674: 165.000 {count=28}
|   |   Distance C ≤ 0.674: 122.000 {count=3}
Distance A ≤ 1.344
|   Distance B > 1.335: 122.000 {count=30}
|   Distance B ≤ 1.335
|   |   Distance A > 1.076
|   |   |   Distance C > 1.798: 79.000 {count=24}
|   |   |   Distance C ≤ 1.798
|   |   |   |   Distance B > 1.112: 165.000 {count=1}
|   |   |   |   Distance B ≤ 1.112: 79.000 {count=14}
|   |   Distance A ≤ 1.076
|   |   |   Distance A > 0.408: 122.000 {count=43}
|   |   |   Distance A ≤ 0.408: 79.000 {count=2}
```

**Figure 34.** Reasoning-based description of the Random Tree shown in Figure 33, that was associated with detecting different values of the X-coordinate of the user based on the associated rules at each node.

Figure 35 shows another of these random trees that was developed by the Random Forest model, as shown in Figure 12, and the reasoning-based description of this tree is shown in Figure 36. This random tree was associated with detecting different values of the Y-coordinate of the user based on the associated rules at each node. We explain the working of the tree for one such detection here, when the user was located at the Y-coordinate—137.000 as per this reference system. This is marked in blue in Figure 35. The comparison starts at the topmost node—Distance A and it is greater than 1.0008 m, so the control moves to the left side of the tree to check Distance B. Here, the value of this distance was greater than 1.334 m, so the control moved to the right side of this node to check for another condition associated with Distance B. Here, it checked if the value of Distance B was greater than 1.286 m or not. For this specific condition as Distance B was greater than 1.286 m so the control traversed to the left side of the node to its child node which is associated with checking another condition at Distance A. This value was less than or equal to 1.095 m, so the Y-coordinate of the user was deduced to be 137.000.

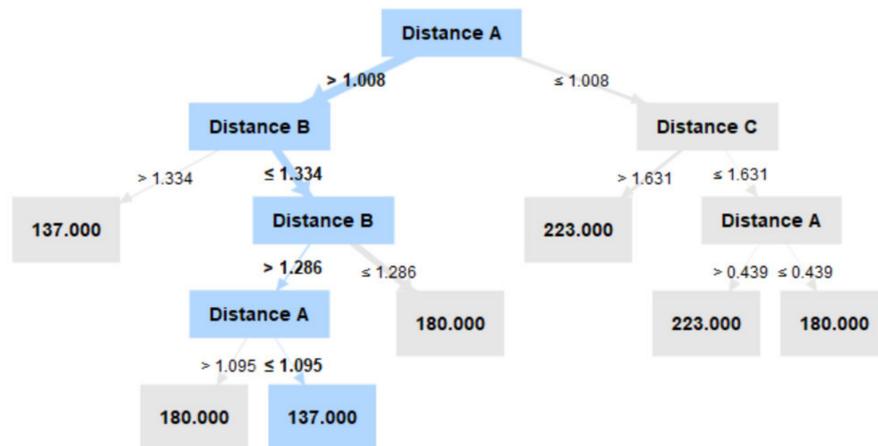

**Figure 35.** Representation of one of the Random Trees developed by the Random Forest-based RapidMiner "process" shown in Figure 12. This random tree was associated with detecting different values of the Y-coordinate of the user based on the associated rules at each node.

```
Distance A > 1.008
|    Distance B > 1.334: 137.000 {count=19}
|    Distance B ≤ 1.334
|    |    Distance B > 1.286
|    |    |    Distance A > 1.095: 180.000 {count=16}
|    |    |    Distance A ≤ 1.095: 137.000 {count=7}
|    |    Distance B ≤ 1.286: 180.000 {count=86}
Distance A ≤ 1.008
|    Distance C > 1.631: 223.000 {count=44}
|    Distance C ≤ 1.631
|    |    Distance A > 0.439: 223.000 {count=2}
|    |    Distance A ≤ 0.439: 180.000 {count=1}
```

**Figure 36.** Reasoning-based description of the Random Tree shown in Figure 35, that was associated with detecting different values of the Y-coordinate of the user based on the associated rules at each node.

Figures 37 and 38 show the output of the RapidMiner "process" (for the first 12 rows), shown in Figure 12, which detected the X-coordinate and Y-coordinate of the user's location based on this methodology, as outlined in Section 4.3. This output was shown by RapidMiner after taking into consideration all the predictions done by each of these 100 Random Trees which were a part of the developed Random Forest-based learning model (Figure 12). The maximum depth of all these random trees was 10. Tables 5 and 6 consist of the description of all the attributes represented in Figures 37 and 38, respectively.

| Row No. | Position X | prediction(Position X) | Distance A | Distance B | Distance C | Time |
|---------|-----------|------------------------|------------|------------|------------|------|
| 1 | 122 | 119.850 | 0.877 | 0.769 | 1.457 | 2017 12:20:22.583 |
| 2 | 122 | 115.980 | 1.202 | 1.031 | 1.893 | 2017 12:20:23.683 |
| 3 | 122 | 128.364 | 1.533 | 1.123 | 2.233 | 2017 12:20:25.915 |
| 4 | 122 | 122.430 | 1.533 | 1.157 | 2.429 | 2017 12:20:30.560 |
| 5 | 122 | 129.238 | 1.533 | 1.174 | 0 | 2017 12:20:35.018 |
| 6 | 122 | 122.860 | 1.473 | 1.206 | 2.735 | 2017 12:20:42.837 |
| 7 | 122 | 122.860 | 1.452 | 1.202 | 2.699 | 2017 12:20:43.958 |
| 8 | 122 | 122 | 1.406 | 1.259 | 2.576 | 2017 12:20:49.533 |
| 9 | 122 | 122 | 1.401 | 1.259 | 2.460 | 2017 12:20:51.772 |
| 10 | 122 | 122 | 1.391 | 1.273 | 2.429 | 2017 12:20:52.880 |
| 11 | 122 | 126.730 | 1.382 | 1.280 | 2.362 | 2017 12:20:54.020 |
| 12 | 122 | 134.470 | 1.366 | 1.285 | 2.362 | 2017 12:20:55.151 |

**Figure 37.** Output (first 12 rows) of the RapidMiner "process" shown in Figure 12 for detection of the spatial coordinates (X-coordinate) of the user in each 'activity-based zone'.

| Row No. | Position Y | prediction(Position Y) | Distance A | Distance B | Distance C | Time |
|---------|-----------|------------------------|------------|------------|------------|------|
| 1 | 180 | 219.990 | 0.877 | 0.769 | 1.457 | 2017 12:20:22.583 |
| 2 | 180 | 191.180 | 1.202 | 1.031 | 1.893 | 2017 12:20:23.683 |
| 3 | 180 | 180 | 1.533 | 1.123 | 2.233 | 2017 12:20:25.915 |
| 4 | 180 | 180 | 1.533 | 1.157 | 2.429 | 2017 12:20:30.560 |
| 5 | 180 | 180 | 1.533 | 1.174 | 0 | 2017 12:20:35.018 |
| 6 | 180 | 180 | 1.473 | 1.206 | 2.735 | 2017 12:20:42.837 |
| 7 | 180 | 180 | 1.452 | 1.202 | 2.699 | 2017 12:20:43.958 |
| 8 | 180 | 179.570 | 1.406 | 1.259 | 2.576 | 2017 12:20:49.533 |
| 9 | 180 | 179.570 | 1.401 | 1.259 | 2.460 | 2017 12:20:51.772 |
| 10 | 180 | 180 | 1.391 | 1.273 | 2.429 | 2017 12:20:52.880 |
| 11 | 180 | 180 | 1.382 | 1.280 | 2.362 | 2017 12:20:54.020 |
| 12 | 180 | 180 | 1.366 | 1.285 | 2.362 | 2017 12:20:55.151 |

**Figure 38.** Output (first 12 rows) of the RapidMiner "process" shown in Figure 12 for detection of the spatial coordinates (Y-coordinate) of the user in each 'activity-based zone'.

**Table 5.** Description of the attributes of the output of the RapidMiner "process" shown in Figure 37.

| Attribute Name | Description |
|----------------|-------------|
| Row No | The row number in the output table |
| Position X | The actual X coordinate of the user's position |
| Prediction (Position X) | The predicted X coordinate of the user's position |
| Distance A | The actual distance of the user from the first Bluetooth beacon |
| Distance B | The actual distance of the user from the second Bluetooth beacon |
| Distance C | The actual distance of the user from the third Bluetooth beacon |
| Time | The associated timestamp information |

**Table 6.** Description of the attributes of the output of the RapidMiner "process" shown in Figure 38.

| Attribute Name | Description |
|----------------|-------------|
| Row No | The row number in the output table |
| Position Y | The actual Y coordinate of the user's position |
| Prediction (Position Y) | The predicted Y coordinate of the user's position |
| Distance A | The actual distance of the user from the first Bluetooth beacon |
| Distance B | The actual distance of the user from the second Bluetooth beacon |
| Distance C | The actual distance of the user from the third Bluetooth beacon |
| Time | The associated timestamp information |

The performance characteristics of this RapidMiner "process", shown in Figure 12, were evaluated by using the RMSE in RapidMiner and the findings are outlined in Table 7. Here, as the predicted attributes were X-coordinate and the Y-coordinate values, both of which were of numerical type, so we were able to use the RMSE method for performance evaluation [68] as recommended by ISO/IEC18305:2016—an international standard for evaluating localization and tracking systems [31]. While RMSE is sometimes calculated by using vector analysis where a single value of RMSE is calculated instead of RMSE along X and Y directions, but as ISO/IEC18305:2016 [31] provides 3 different formulae in Chapter 8, for calculation of RMSE in X-direction, Y-direction, and the associated Horizontal Error, so we calculated these performance metrics separately. The formulae for calculation of these three performance characteristics, as mentioned in ISO/IEC18305:2016, are represented in Equations (1)–(3).

$$\varepsilon_{x,rms} = \sqrt{\frac{1}{N} \sum_{i=1}^{N} \varepsilon_{x,i}^2} \qquad (1)$$

$$\varepsilon_{y,rms} = \sqrt{\frac{1}{N} \sum_{i=1}^{N} \varepsilon_{y,i}^2} \qquad (2)$$

$$\varepsilon_{h,rms} = \sqrt{\varepsilon_{x,rms}^2 + \varepsilon_{y,rms}^2} \qquad (3)$$

where:

$\varepsilon_{x,rms}$ stands for RMSE in the X-direction

$\varepsilon_{y,rms}$ stands for RMSE in the Y-direction

$\varepsilon_{h,rms}$ stands for Horizontal Error that considers RMSE in the X-direction and RMSE in the Y-direction

$\varepsilon_{x,i}^2$ stands for squared errors in the X-direction

$\varepsilon_{y,i}^2$ stands for squared errors in the Y-direction

N stands for sample size

**Table 7.** Description of the performance characteristics of the Random Forest-based RapidMiner "process" shown in Figure 12.

| Description of Performance Characteristic | Value |
|---|---|
| Root Mean Squared Error for detection of X-coordinate | 5.85 cm |
| Root Mean Squared Error for detection of Y-coordinate | 5.36 cm |
| Horizontal Error | 7.93 cm |

As can be observed from Table 7, the root mean squared error for detection of the instantaneous X-coordinate and Y-coordinate values of the user's position were found to be 5.85 cm and 5.36 cm, respectively. To add to the above, the Horizontal Error was found to be 7.93 cm. Further discussion about how this approach and the associated results and findings address multiple research challenges in this field is presented in Section 7.

## 6. Deducing the Optimal Machine Learning Model for Indoor Localization

As outlined in Section 2, one the research challenges in this field of Indoor Localization is the need to develop an optimal machine learning model for Indoor Localization systems, Indoor Positioning Systems, and Location-Based Services. In [9–25], researchers have used multiple machine learning approaches—Random Forest, Artificial Neural Network, Decision Tree, Support Vector Machine, k-NN, Gradient Boosted Trees, Deep Learning, and Linear Regression. However, none of these works implemented multiple machine learning models to evaluate and compare the associated performance characteristics to deduce the optimal machine learning approach. Due to the differences in the datasets used or the real-time data that was collected, the associated data preprocessing steps that were different, variations in train and test ratio of the data, and several other dissimilar steps that were associated with the developments of each of these machine learning models as presented in [9–25], their final performance accuracies cannot be directly compared to deduce the best approach. Thus, analyzing the performance characteristics of multiple machine learning models, developed, implemented, and tested as per the same methodology, to deduce the optimal approach for development of such Indoor Localization systems serves as the main motivation for the work presented in the section.

In Section 4.3, we outlined the steps associated with the proposed methodology to detect the spatial coordinates of a user's indoor position based on the associated user interactions with the context parameters and the user-centered local spatial context, by using a reference system. Upon development of the same, as a RapidMiner "process", as shown in Figure 12, by using the Random Forest-based learning approach, we evaluated its performance characteristics by calculating the RMSE for X coordinate, the RMSE for Y coordinate, and the Horizontal Error—these are performance evaluation metrics mentioned in ISO/IEC18305:2016 [31]. The RMSE for detection of the X-coordinate and Y-coordinate were found to be 5.85 cm and 5.36 cm as per Equations (1) and (2), respectively. The associated Horizontal Error as per Equation (3) was found to be 7.93 cm. These results are shown in Table 7. In this section, we followed the same steps as outlined in Section 4.3 and as per the flowchart shown in Figure 11, to develop, implement, and test this methodology by using all the machine learning methods that have been used by researchers [9–25] in this field. These machine learning methods included—Random Forest, Artificial Neural Network, Decision Tree, Support Vector Machine, k-NN, Gradient Boosted Trees, Deep Learning, and Linear Regression. As we had already developed this approach by using the Random Forest approach (Figure 12 and Section 5.3), so, we did not repeat the same in this section and we performed this study on all the other machine learning methods. For each of these methods we developed a RapidMiner "process" by using the same version of RapidMiner on the same computer as outlined in Section 3. Each of these processes were developed as a combination of built-in "operators" and user defined "operators" in RapidMiner. An overview of built-in "operators" and user-defined "operators" in RapidMiner was presented in Section 3. We used the same system architecture as outlined in Figure 11 for development of all the machine learning-based "processes" in RapidMiner as discussed here, so, a separate system architecture is not provided in this section. The specific steps that we followed for development of each of these RapidMiner processes corresponding to these different machine learning methods are as follows:

i.   Use the 'Dataset' "operator" to import the dataset [67] into the RapidMiner "process".

ii.  Utilize the 'Data-Preprocess' "operator" to perform multiple preprocessing steps (Steps iii to vi in Section 4.3) prior to splitting the data for training and testing. We developed this 'Data-Preprocess' "operator".

iii. Use the built-in "operator" called 'Split Data' to divide the dataset into training set and test set. The dataset [67] consisted of 250 rows. We used 70% of the data for training and the remaining 30% for testing.

iv.  Use the specific machine learning model to train the system. By specific machine learning model, we mean either the usage of the Artificial Neural Network or Decision Tree or Support Vector Machine or k-NN or Gradient Boosted Trees or Deep Learning or Linear Regression. These machine learning models are present in RapidMiner as built-in "operators" that can be directly used. However, a few of these learning models in RapidMiner such as Artificial Neural Network, Support Vector Machines, and Linear Regression sometimes need the 'nominal to numerical' "operator" for training and testing of the model, based on the characteristics and nature of the dataset being used.

v.   Utilize the built-in 'Apply Model' "operator" to apply the learning model on the test data. This "operator" was renamed in each of these "processes", as per the specific learning model that was being developed and evaluated, to indicate the differences in the associated functionalities of this "operator" for each of these RapidMiner "processes". For the RapidMiner "process" that used the Artificial Neural Network, the 'Apply Model' "operator" was renamed to 'Apply Model—ANN'. Similarly, for the machine learning models—Decision Tree, Support Vector Machine, k-NN, Gradient Boosted Trees, Deep Learning, and Linear Regression, this "operator" was renamed to 'Apply Model—DT', 'Apply Model—SVM', 'Apply Model—kNN', 'Apply Model—GBT', 'Apply Model—DL', and 'Apply Model—LR', respectively.

vi. Use the built-in 'Performance' "operator" to evaluate the performance characteristics of the "process" by calculating the RMSE in X-direction, the RMSE in Y-direction, and the Horizontal Error as per Equations (1)–(3), respectively. This "operator" was renamed in each of these "processes", as per the specific learning model that was being developed and evaluated, to indicate the differences in the associated functionalities of this "operator" for each of these RapidMiner "processes". For the RapidMiner "process" that used the Artificial Neural Network, the 'Performance' "operator" was renamed to 'ANN-Performance'. Similarly, for the machine learning models—Decision Tree, Support Vector Machine, k-NN, Gradient Boosted Trees, Deep Learning, and Linear Regression, this "operator" was renamed to 'DT-Performance', 'SVM-Performance', 'kNN-Performance', 'GBT-Performance', 'DL-Performance', and 'LR-Performance', respectively.

These RapidMiner "processes", that were developed by using the learning approaches—Artificial Neural Network, Decision Tree, Support Vector Machine, k-NN, Gradient Boosted Trees, Deep Learning, and Linear Regression, are shown in Figures 39–45, respectively. The corresponding performance metrics in terms of the RMSE in X-direction, the RMSE in Y-direction, and the Horizontal Error are shown in Tables 8–14, respectively. We did not develop the RapidMiner "process" by using the Random Forest approach in this section as we had already developed the same in Figure 12 and discussed its performance characteristics in terms of the RMSE in X-direction, the RMSE in Y-direction, and the Horizontal Error in Table 7.

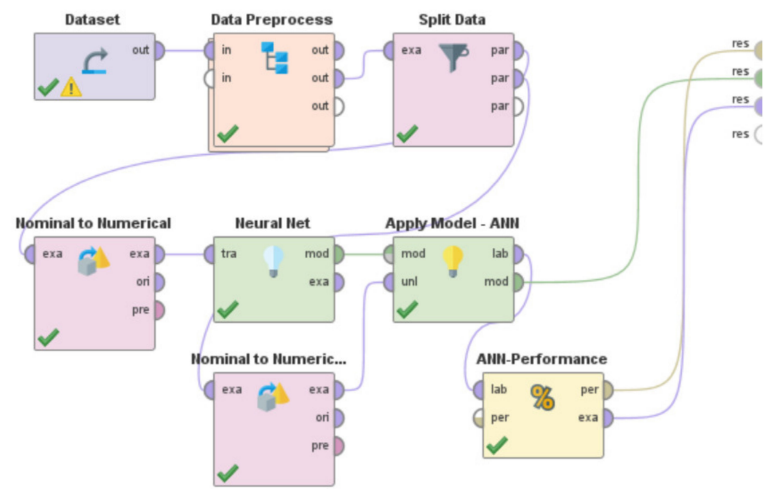

**Figure 39.** "Process" developed in RapidMiner that used an Artificial Neural Network (ANN)-based learning approach and followed the steps outlined in Section 4.3 for detection of the spatial coordinates of the user in each 'activity-based zone'.

**Table 8.** Description of the performance characteristics of the Artificial Neural Network (ANN)-based RapidMiner "process" shown in Figure 39.

| Description of Performance Characteristic | Value |
|---|---|
| Root Mean Squared Error for detection of X-coordinate | 28.00 cm |
| Root Mean Squared Error for detection of Y-coordinate | 16.16 cm |
| Horizontal Error | 32.33 cm |

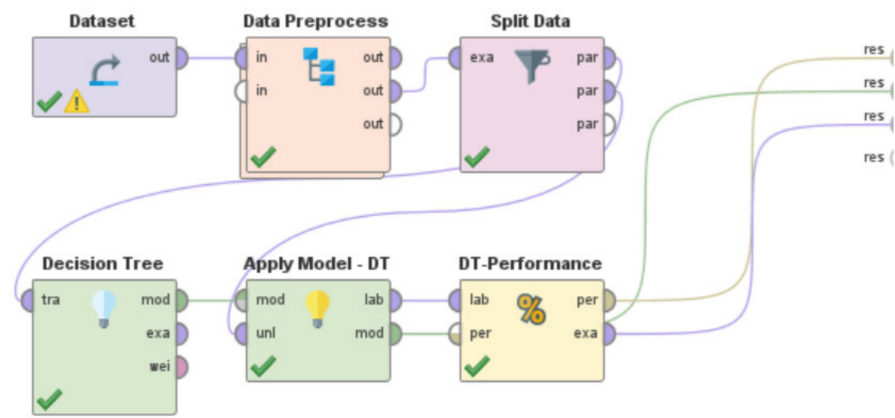

**Figure 40.** "Process" developed in RapidMiner that used a Decision Tree (DT)-based learning approach and followed the steps outlined in Section 4.3 for detection of the spatial coordinates of the user in each 'activity-based zone'.

**Table 9.** Description of the performance characteristics of the Decision Tree (DT)-based RapidMiner "process" shown in Figure 40.

| Description of Performance Characteristic | Value |
|---|---|
| Root Mean Squared Error for detection of X-coordinate | 12.52 cm |
| Root Mean Squared Error for detection of Y-coordinate | 6.19 cm |
| Horizontal Error | 13.97 cm |

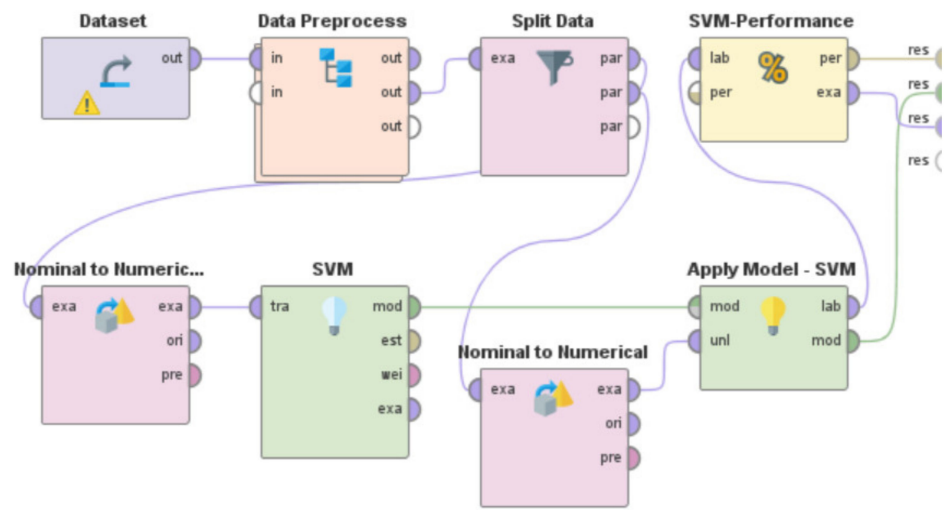

**Figure 41.** "Process" developed in RapidMiner that used a Support Vector Machine (SVM)-based learning approach and followed the steps outlined in Section 4.3 for detection of the spatial coordinates of the user in each 'activity-based zone'.

**Table 10.** Description of the performance characteristics of the Support Vector Machine (SVM)-based RapidMiner "process" shown in Figure 41.

| Description of Performance Characteristic | Value |
|---|---|
| Root Mean Squared Error for detection of X-coordinate | 27.92 cm |
| Root Mean Squared Error for detection of Y-coordinate | 27.17 cm |
| Horizontal Error | 38.96 cm |

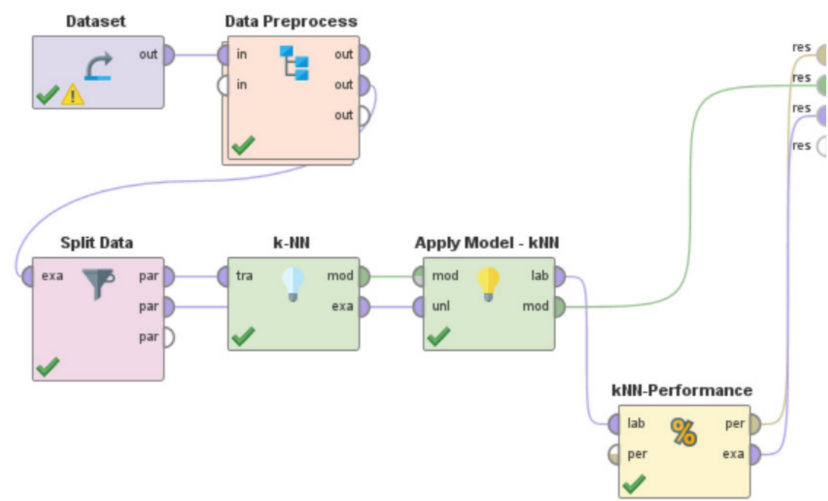

**Figure 42.** "Process" developed in RapidMiner that used a kNN-based learning approach and followed the steps outlined in Section 4.3 for detection of the spatial coordinates of the user in each 'activity-based zone'.

**Table 11.** Description of the performance characteristics of the kNN-based RapidMiner "process" shown in Figure 42.

| Description of Performance Characteristic | Value |
| --- | --- |
| Root Mean Squared Error for detection of X-coordinate | 10.11 cm |
| Root Mean Squared Error for detection of Y-coordinate | 2.96 cm |
| Horizontal Error | 10.54 cm |

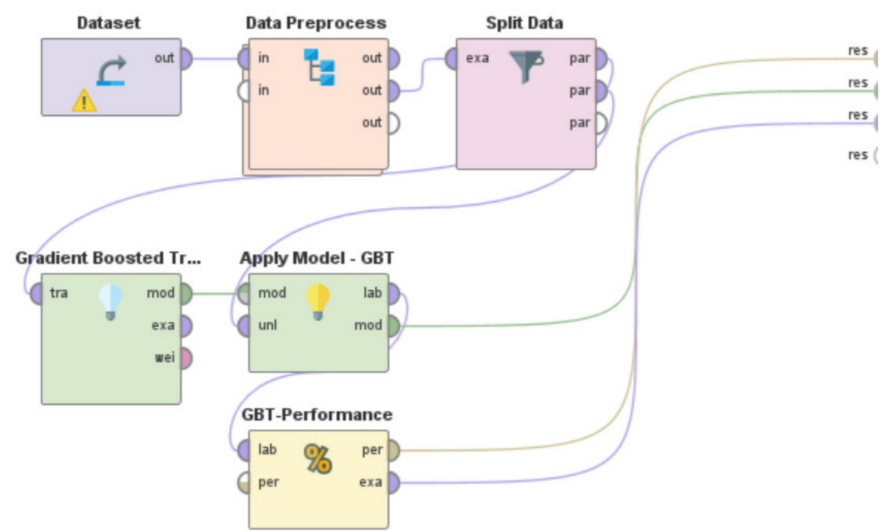

**Figure 43.** "Process" developed in RapidMiner that used a Gradient Boosted Trees (GBT)-based learning approach and followed the steps outlined in Section 4.3 for detection of the spatial coordinates of the user in each 'activity-based zone'.

**Table 12.** Description of the performance characteristics of the Gradient Boosted Trees (GBT)-based RapidMiner "process" shown in Figure 43.

| Description of Performance Characteristic | Value |
| --- | --- |
| Root Mean Squared Error for detection of X-coordinate | 28.12 cm |
| Root Mean Squared Error for detection of Y-coordinate | 27.65 cm |
| Horizontal Error | 39.44 cm |

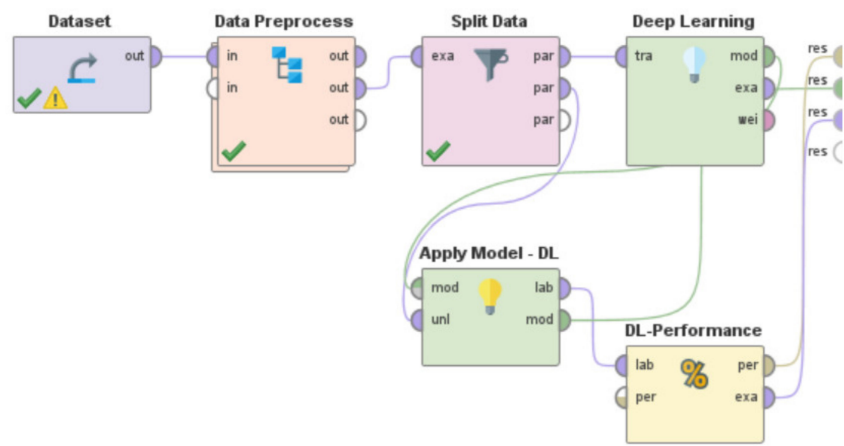

**Figure 44.** "Process" developed in RapidMiner that used a Deep Learning (DL)-based learning approach and followed the steps outlined in Section 4.3 for detection of the spatial coordinates of the user in each 'activity-based zone'.

**Table 13.** Description of the performance characteristics of the Deep Learning (DL)-based RapidMiner "process" shown in Figure 44.

| Description of Performance Characteristic | Value |
|---|---|
| Root Mean Squared Error for detection of X-coordinate | 29.67 cm |
| Root Mean Squared Error for detection of Y-coordinate | 12.04 cm |
| Horizontal Error | 32.02 cm |

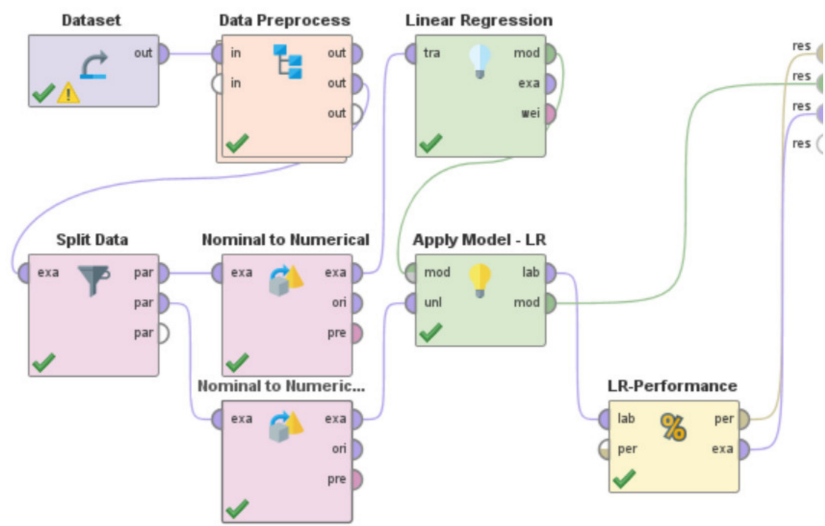

**Figure 45.** "Process" developed in RapidMiner that used a Linear Regression (LR)-based learning approach and followed the steps outlined in Section 4.3 for detection of the spatial coordinates of the user in each 'activity-based zone'.

**Table 14.** Description of the performance characteristics of the Linear Regression (LR)-based RapidMiner "process" shown in Figure 45.

| Description of Performance Characteristic | Value |
|---|---|
| Root Mean Squared Error for detection of X-coordinate | 28.064 cm |
| Root Mean Squared Error for detection of Y-coordinate | 27.630 cm |
| Horizontal Error | 39.382 cm |

The performance metrics—RMSE in X-direction, RMSE in Y-direction, and Horizontal Error for all these machine learning models—Random Forest, Artificial Neural Network, Decision Tree, Support Vector Machine, k-NN, Gradient Boosted Trees, Deep Learning, and Linear Regression are summarized in Table 15. The analysis of these metrics is shown in Figure 46.

**Table 15.** Comparison of the performance metrics of the different learning approaches—Random Forest, Artificial Neural Network, Decision Tree, Support Vector Machine, k-NN, Gradient Boosted Trees, Deep Learning, and Linear Regression.

| Learning Approach | Performance Metrics | | |
| --- | --- | --- | --- |
| | RMSE in X-Direction | RMSE in Y-Direction | Horizontal Error |
| Random Forest | 5.85 cm | 5.36 cm | 7.93 cm |
| Artificial Neural Network | 28.00 cm | 16.16 cm | 32.33 cm |
| Decision Tree | 12.52 cm | 6.19 cm | 13.97 cm |
| Support Vector Machine | 27.92 cm | 27.17 cm | 38.96 cm |
| k-NN | 10.11 cm | 2.96 cm | 10.54 cm |
| Gradient Boosted Trees | 28.12 cm | 27.65 cm | 39.44 cm |
| Deep Learning | 29.67 cm | 12.04 cm | 32.02 cm |
| Linear Regression | 28.06 cm | 27.63 cm | 39.38 cm |

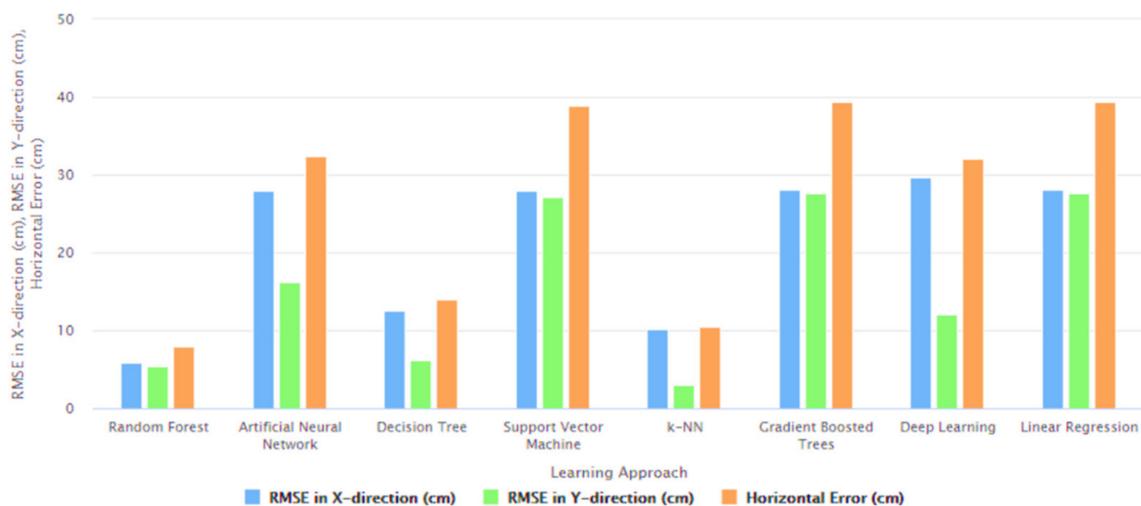

**Figure 46.** Comparison of the performance metrics of the different learning approaches—Random Forest, Artificial Neural Network, Decision Tree, Support Vector Machine, k-NN, Gradient Boosted Trees, Deep Learning, and Linear Regression, shown in the form a Bar (Column) Style Plot.

From Table 15 and Figure 46 the following can be observed and deduced:

i.    The Random Forest-based learning approach has the least Horizontal Error of 7.93 cm and the Gradient Boosted Trees-based learning approach has the highest Horizontal Error of 39.44 cm. Considering Horizontal Error as a function, where Horizontal Error (x) gives the Horizontal Error of 'x', where 'x' is a machine learning model; the Horizontal Errors of these machine learning models can be arranged in an increasing to decreasing order as: Horizontal Error (Random Forest) < Horizontal Error (k-NN) < Horizontal Error (Decision Tree) < Horizontal Error (Deep Learning) < Horizontal Error (Artificial Neural Network) < Horizontal Error (Support Vector Machine) < Horizontal Error (Linear Regression) < Horizontal Error (Gradient Boosted Trees).

ii.   The RMSE in X-direction is least for the Random Forest-based learning approach and is highest for the Deep Learning-based learning approach with the respective values being 5.85 cm and 29.67 cm, respectively. Considering RMSE in X-direction as a function, where RMSE in X-direction (p) gives the RMSE in X-direction of 'p', where 'p' is a machine learning model; the RMSE in X-directions of these machine learning models can be arranged in an increasing to decreasing order as: RMSE in X-direction (Random Forest) < RMSE in X-direction (k-NN) < RMSE in X-direction (Decision Tree) < RMSE in X-direction (Support Vector Machine) < RMSE in X-direction (Artificial Neural Network) < RMSE in X-direction (Linear Regression) < RMSE in X-direction (Gradient Boosted Trees) < RMSE in X-direction (Deep Learning).

iii. The RMSE in Y-direction is least for the k-NN-based learning approach with a value of 2.96 cm and this metric is highest for the Gradient Boosted Trees-based learning approach with a value of 27.65 cm. Considering RMSE in Y-direction as a function, where RMSE in Y-direction (q) gives the RMSE in Y-direction of 'q', where 'q' is a machine learning model; the RMSE in Y-directions of these machine learning models can be arranged in an increasing to decreasing order as: RMSE in Y-direction (k-NN) < RMSE in Y-direction (Random Forest) < RMSE in Y-direction (Decision Tree) < RMSE in Y-direction (Deep Learning) < RMSE in Y-direction (Artificial Neural Network) < RMSE in Y-direction (Support Vector Machine) < RMSE in Y-direction (Linear Regression) RMSE in Y-direction (Gradient Boosted Trees)

As can be seen from (i) and (ii) above, the Random Forest-based learning approach has the least RMSE in X-direction as well as the least Horizontal Error. The respective values being 5.85 cm and 7.93 cm, respectively. Even though the k-NN based learning approach has a lesser RMSE in Y-direction (2.96 cm) as compared to RMSE of Random Forest in Y-direction (5.36 cm), the overall Horizontal Error for the k-NN based learning approach is much higher as compared to the Horizontal Error of the Random Forest-based learning approach with the respective values being 10.54 cm and 7.93 cm. Thus, for all practical purposes it may be concluded that the Random Forest-based learning approach is the optimal machine learning model for development of Indoor Localization systems, Indoor Positioning Systems, and Location-Based Services. Further discussion about how this comparative study and the associated results and findings address multiple research challenges in this field is presented in Section 7.

## 7. Comparative Discussion

Despite several advances in the fields of Indoor Localization, Indoor Positioning Systems, Human Activity Recognition, Activity Analysis, and Ambient Assisted Living, there exist several research challenges in this field. The work presented in this paper at the intersection of Big Data, Machine Learning, Indoor Localization, Ambient Assisted Living, Internet of Things, Activity Centric Computing, Human–Computer Interaction, Pattern Recognition, and Assisted Living Technologies, and their related application domains aims to take a comprehensive approach to address these challenges. We introduced these research challenges in Section 2. In this section, we further discuss the same and outline how the work presented in this paper and the associated results and findings addresses these challenges and outperform similar works in this field. This is discussed as follows:

1. Need for AAL-based activity recognition and activity analysis-based systems to be able to track the indoor location of the user: The AAL-based systems currently lack the ability to track the indoor location of the user. There have been several works done [46–51] in these interrelated fields of activity recognition, activity analysis, and fall detection, but none of these works have focused on Indoor Localization. Being able to track the indoor location of a user is of prime importance and of crucial need for AAL-based systems to be able to contribute towards improving the quality of life of individuals in the future of living environments, such as, Smart Homes. For instance, an elderly person could be staying in an apartment which is a part of a multistoried building such as Taipei 101 [70] or Burj Khalifa [71]—both of which are amongst the tallest buildings in the world. When this elderly person experiences a fall, a fall detection system such as [51], could detect a fall and alert caregivers but the current GPS-based technologies would only provide the building level information.

The lack of the precise location information in terms of the specific floor, apartment, and room, could cause delay of medical attention or assistive care. Such delay of care can have both short-term and long-term health-related impacts to the elderly such as long lie [72], that can cause dehydration, rhabdomyolysis, pressure injuries, carpet burns, hypothermia, pneumonia, and fear of falling, which could lead to decreased independence and willingness in carrying out daily routine activities. Long lie can even lead to death in some cases. Thus, it is the need of the hour that AAL-based systems should not only be able to track, monitor, and analyze human behavior but they should also be equipped with the functionality to detect the indoor location of the users. The work presented in this paper addresses this challenge by proposing a novel Big-Data driven methodology that can study the multimodal components of user interactions during Activities of Daily Living (ADLs) (Tables 1 and 2) and analyze the data from BLE beacons and BLE scanners to track a user's indoor location in a specific 'activity-based zone' during different ADLs (Figure 6). This approach was developed by using a k-nearest neighbor (k-NN)-based learning approach (Section 4.1). When tested on a dataset (Figure 17, Table 3) it achieved a performance accuracy of 81.36% (Figures 18 and 19).

2. Need for context-independent Indoor Localization systems: As outlined in Section 2, several recent works related to Indoor Localization systems are context-based and are only functional in the specific environments for which they were developed [26–30]. These specific environments include—factories [26], indoor parking [27], hospitals [28], industry-based settings [29], and academic environments [30]. For instance, the methodology proposed in [29] is not functional in any of the settings described in [26–28,30]. The future of interconnected Smart Cities would consist of a host of indoor environments in the living and functional spaces of humans, which would be far more diverse, different, and complicated as compared to the environments described in [26–30]. The challenge is thus to develop a means for Indoor Localization that is not environment dependent and can be seamlessly deployed in any IoT-based setting irrespective of the associated context parameters and their attributes. The work proposed in this paper addresses this challenge by proposing a novel context independent approach that can interpret the accelerometer and gyroscope data from diverse behavioral patterns to detect the 'zone-based' indoor location of a user in any IoT-based environment (Section 4.2). This proposed approach (Figure 9) can study, analyze, and interpret the distinct behavioral patterns, in terms of the associated accelerometer and gyroscope data, local to each such 'zone', in the confines of any given IoT-based space without being affected by the changes or variations in the context parameters or environment variables. It uses a Random Forest-based learning approach for the training and the same was evaluated on a dataset (Figure 26, Table 4). The performance accuracy of this method for detecting a user's location in each of these 'zones', that were present in this dataset [66], was found to be 81.13% (Figures 27 and 28). Here, the 'zone-based' mapping of a user's location refers to mapping the user in one of the multiple 'activity-based zones' that any given IoT-based environment can be classified into based on the specific activity being performed by the user. The accelerometer and gyroscope data are user behavior dependent and not context parameter dependent and neither is this approach of spatially mapping a given IoT-based space into 'activity-bases zones' dependent on any specific set of context parameters, as explained in Section 4.2. This upholds the context independent nature of this methodology. In other words, this proposed methodology can be seamlessly applied to any IoT-based environment, including all the environments described in [26–30], as well as in any other IoT-based setting that involves different forms of user interactions on context parameters or environment variables, which can be characterized by the changes in the associated behavioral data.

3. The RMSE of the existing Indoor Localization systems [33–43] are still high and greater precision and accuracy for detection of indoor location is the need of the hour. Several performance metrics have been used by researchers for studying the characteristics of Indoor Localization systems, Indoor Positioning Systems, and Location-Based Services. However, ISO/IEC18305:2016, an international standard for evaluating localization and tracking systems [31], which is one of the recent works in this field, lists several metrics and the associated formulae for evaluating the performance characteristics of such systems. These include the formulae for determination of the RMSE in the X-direction, Y-direction, and in the X-Y plane. When the RMSE is determined in the X-Y plane, it is referred to as Horizontal Error as per the definitions of the standard [31]. We have presented and discussed the associated formulae in Equations (1)–(3). Upon reviewing the recent works [33–43] related to this field, as presented in Section 2, it can be observed that the RMSE of the works are still significantly high in view of the average dimensions of an individual's living space. As per [44,45], (1) the average dimensions of newly built one-bedroom apartments and two-bedroom apartments in United States in 2018 were 757 square feet (70.3276 square meters) and 1138 square feet (105.7236 square meters), respectively. In view of these dimensions of these apartments, it can be concluded that higher precision is needed for the future of Indoor Localization systems. Such systems should have much lower values of RMSE in X and Y directions as well as their overall Horizontal Error should be low. The work presented in this paper addresses this research challenge by proposing a methodology to detect the spatial coordinates of a user's indoor position based on the associated user interactions with the context parameters and the user-centered local spatial context, by using a reference system. In Section 4.3 we have presented the steps for development of this approach for Indoor Localization and the results of the same are discussed in Section 5.3. While RMSE is sometimes calculated by using vector analysis where a single value of RMSE is calculated instead of RMSE along X and Y directions, but as ISO/IEC18305:2016 [31] presents two separate formulae (Equations (1) and (2)) for calculation of RMSE in X-direction and RMSE in Y-direction and a third formula (Equation (3)) for Horizontal Error calculation, so, we calculated RMSE in X and Y directions separately and then calculated the Horizontal Error as per Equations (1)–(3), respectively. As can be seen from the results (Table 7), the performance characteristics of our approach are—RMSE in X-direction: 5.85 cm, RMSE in Y-direction: 5.36 cm, and Horizontal Error: 7.93 cm. As can be seen from [33–43], RMSE is usually represented in meters, so upon converting these metrics from Table 7 to meters (correct to 2 decimal places) the corresponding values are: RMSE in X-direction: 0.06 m, RMSE in Y-direction: 0.05 m, and Horizontal Error: 0.08 m. The RMSE of these existing works [33–43], in increasing to decreasing order are shown in Table 16.

**Table 16.** Summary of the various Indoor Localization approaches that used RMSE for evaluation of the performance metrics.

| RMSE Value (in Meters) | Work(s) |
| --- | --- |
| 0.32 | Bolic et al. [34] |
| 1.00 | Chen et al. [41] |
| 1 to 2 | Angermann et al. [35] |
| 1.20 | Klingbeil et al. [38] |
| 1.28 | Chen at al. [43] |
| 1.40 | Correa et al. [33] |
| 1.53 | Evennou et al. [36] |
| 2.90 | Li et al. [42] |
| 3.10 | Liu et al. [40] |
| 4.30 | Wang et al. [37] |
| 4.55 | Pei et al. [39] |

From Table 16, it can be concluded that Bolic et al.'s work [34] has the best performance accuracy out of all the works reviewed in [33–43] with the RMSE being 0.32 m. Upon comparing the performance metrics of our approach (Table 7) with Bolic et al.'s work [34], it can be easily concluded that our work outperforms the same in terms of performance accuracy as the RMSE values (RMSE in X-direction: 0.06 m, RMSE in Y-direction: 0.05 m, and Horizontal Error: 0.08 m) of our methodology are significantly lower. As our work outperforms Bolic et al.'s work, which has the best accuracy out of all the works reviewed in [33–43], so, it can also be concluded that our work outperforms all the other works as well [33,35–43], in terms of the RMSE method of performance evaluation, as recommended by ISO/IEC18305:2016 [31].

4.  Need for an optimal machine learning-based approach for Indoor Localization: A range of machine learning approaches—Random Forest, Artificial Neural Network, Decision Tree, Support Vector Machine, k-NN, Gradient Boosted Trees, Deep Learning, and Linear Regression, have been used by several researchers [9–25] for development of various types of Indoor Localization systems for IoT-based environments. While each of these systems seem to perform reasonably well but none of these works attempted to develop an optimal machine learning model for Indoor Localization systems. Additionally, due to variations in the data source, differences in the types of data, varied methods of data collection, different training set to test set ratios, dissimilar data preprocessing steps, as well as because of differences in the simulated or real-world environments in which these respective systems were developed, implemented, and deployed, the performance metrics of these systems cannot be directly compared to deduce the optimal approach. These works [9–25] along with the machine learning approaches that were used in each are outlined in Table 17.

**Table 17.** Summary of the various machine learning approaches that have been investigated by researchers in this field.

| Learning Approach Used | Work(s) |
| --- | --- |
| Random Forest | Varma et al. [19], Gao et al. [20] |
| Artificial Neural Network | Khan et al. [16], Labinghisa et al. [17], Qin et al. [18] |
| Decision Tree | Musa et al. [9], Yim et al. [10] |
| Support Vector Machine | Sjoberg et al. [11], Zhang et al. [12] |
| k-NN | Zhang et al. [13], Ge et al. [14], Hu et al. [15] |
| Gradient Boosted Trees | Wang et al. [25] |
| Deep Learning | Zhang et al. [23], Poulose et al. [24] |
| Linear Regression | Jamâa et al. [21], Barsocchi et al. [22] |

There is a need to address this research challenge of identifying the optimal machine learning methodology for Indoor Localization. The work presented in this paper addresses this challenge. In Section 6—we developed, implemented, and tested the performance characteristics of different learning models to perform Indoor Localization by using the same dataset [67], the same data preprocessing steps, the same training and test ratios, and the same methodology, which we presented in Section 4.3. The learning models that we developed and studied included—Random Forest, Artificial Neural Network, Decision Tree, Support Vector Machine, k-NN, Gradient Boosted Trees, Deep Learning, and Linear Regression. These models were developed to detect the spatial coordinates of a user's indoor location as per the methodology outlined in Section 4.3. RapidMiner was used to develop these machine learning models, and the corresponding RapidMiner "processes" are shown in Figures 12 and 39–45, respectively. We evaluated the performance characteristics of these models based on three performance metrics as outlined in ISO/IEC18305:2016—an international standard for evaluating localization and tracking systems [31]. These include—RMSE in X-direction, RMSE in Y-direction, and Horizontal Error (Equations (1)–(3)). The performance characteristics of these respective machine learning models are shown in Tables 7–14. In Table 15 and Figure 46, we present the comparisons amongst these learning models to deduce the optimal machine learning approach for development of an Indoor Localization system. Based on the findings presented in Table 15 and Figure 46, the following can be observed:

i. Out of all these learning approaches, the Random Forest-based learning approach has the least Horizontal Error of 7.93 cm. In an increasing to decreasing order, the Horizontal Errors of these machine learning models can be arranged as: Horizontal Error (Random Forest) < Horizontal Error (k-NN) < Horizontal Error (Decision Tree) < Horizontal Error (Deep Learning) < Horizontal Error (Artificial Neural Network) < Horizontal Error (Support Vector Machine) < Horizontal Error (Linear Regression) < Horizontal Error (Gradient Boosted Trees).

ii. Out of all these learning approaches, the RMSE in X-direction is the least for the Random Forest-based learning approach, which is equal to 5.85 cm. In an increasing to decreasing order, the RMSE in X-direction of these machine learning models can be arranged as: RMSE in X-direction (Random Forest) < RMSE in X-direction (k-NN) < RMSE in X-direction (Decision Tree) < RMSE in X-direction (Support Vector Machine) < RMSE in X-direction (Artificial Neural Network) < RMSE in X-direction (Linear Regression) < RMSE in X-direction (Gradient Boosted Trees) < RMSE in X-direction (Deep Learning).

iii. Out of all these learning models, the RMSE in Y-direction of the k-NN-based learning approach is the lowest and the RMSE in Y-direction of the Random Forest-based learning approach is the second lowest. Their respective values being 2.96 cm and 5.36 cm, respectively. In an increasing to decreasing order, the RMSE in Y-direction of these machine learning models can be arranged as: RMSE in Y-direction (k-NN) < RMSE in Y-direction (Random Forest) < RMSE in Y-direction (Decision Tree) < RMSE in Y-direction (Deep Learning) < RMSE in Y-direction (Artificial Neural Network) < RMSE in Y-direction (Support Vector Machine) < RMSE in Y-direction (Linear Regression) RMSE in Y-direction (Gradient Boosted Trees)

iv. From (i) and (ii), it can be deduced that for the RMSE in X-direction and for the Horizontal Error (Equations (1) and (3)) methods of performance evaluation, the Random Forest-based learning approach outperforms all the other learning approaches—Artificial Neural Network, Decision Tree, Support Vector Machine, k-NN, Gradient Boosted Trees, Deep Learning, and Linear Regression. Even though the k-NN-based learning approach performs better than the Random Forest-based learning approach for determination of the RMSE in Y-direction, as can be seen from (iii), however, the difference between RMSE in Y-direction for the k-NN based learning approach and the RMSE in Y-direction for the Random Forest based learning approach is not high. To add, for the other two performance metrics—RMSE in X-direction and Horizontal Error, the k-NN based learning approach does not perform as good as the Random Forest-based learning approach and its error values are much higher. Thus, based on these findings and the discussions, which are presented in an elaborate manner in Section 6, it can be concluded that a Random Forest-based learning approach is the optimal machine learning model for development of Indoor Localization systems, Indoor Positioning Systems, and Location-Based Services.

## 8. Conclusions and Scope for Future Work

The future of technology-laden living and functional environments, for instance, Smart Homes, Smart Cities, Smart Workplaces, Smart Industries, and Smart Vehicles, would involve Human–Computer, Human–Robot, Human–Machine, and other forms of human interactions with technology-laden gadgets, systems, or devices. Although Global Positioning System (GPS) and Global Navigation Satellite Systems (GNSS) have significantly revolutionized navigation research by being able to track people, objects, and assets in real-time, such technologies are still ineffective in indoor settings [1]. Indoor Localization has multiple applications in the context of such forms of human interactions with technology. As per [3], the market opportunities of Indoor Localization related systems are expected to be in the order of USD 10 billion by 2024 due to the diverse societal needs that such systems can address. There can be multiple use cases and applications of Indoor Localization systems that can be investigated and studied. This paper focuses on one specific application domain—Ambient Assisted Living (AAL) of elderly people in the future of Internet of Things (IoT)-based living environments, such as Smart Homes and Smart Cities. The work presented in this paper addresses multiple research challenges and makes several scientific contributions to this field by integrating the latest advancements from Big Data, Machine Learning, Indoor Localization, Ambient Assisted Living, Internet of Things, Activity Centric Computing, Human–Computer Interaction, Pattern Recognition, Assisted Living Technologies, and their related application domains.

First, to address the research challenge that the AAL-based systems and technologies [46–51] for activity recognition, activity analysis, and fall detection, currently lack the ability to track the indoor location of the user; this paper proposes a novel Big-Data driven methodology that studies the multimodal components of user interactions and analyzes the data from BLE beacons and BLE scanners to track a user's indoor location in a specific 'activity-based zone' during Activities of Daily Living. This approach was developed by using a k-nearest neighbor (k-NN)-based learning approach. When tested on a dataset this methodology achieved a performance accuracy of 81.36%.

Second, to address the limitation in several Indoor Localization systems [26–30], that they are context-based and are only functional in the specific environments in which they were developed; this paper proposes a context independent approach that can interpret the accelerometer and gyroscope data from diverse behavioral patterns to detect the 'zone-based' indoor location of a user in any IoT-based environment. Here, the 'zone-based' mapping of a user's location refers to mapping the user in one of the multiple 'activity-based zones' that any given IoT-based environment can be classified into, based on the specific activity being performed by the user. This methodology was developed by using the Random Forest-based learning approach. When tested on a dataset this novel methodology achieved a performance accuracy of 81.13%.

Third, to address the challenge that the RMSE of the existing Indoor Localization systems are still high [33–43] and greater precision and accuracy for detection of indoor location is the need of the hour; this paper proposes a methodology to detect the spatial coordinates of a user's indoor position based on the associated user interactions with the context parameters and the user-centered local spatial context, by using a reference system. The performance characteristics of this system were evaluated as per three metrics stated in ISO/IEC18305:2016 [31], which is an international standard for testing Indoor Localization and Tracking Systems. These metrics included root mean squared error (RMSE) in X-direction, RMSE in Y-direction, and the Horizontal Error which were found to be 5.85 cm, 5.36 cm, and 7.93 cm, respectively. A comparison study of this approach with similar researches [33–43] in this field showed that our system outperformed all these works that had used a similar approach of performance evaluation.

Finally, in view of the fact that multiple machine learning-based approaches have been used by researchers [9–25] and there is a need to identify the optimal machine learning model that can be used to develop the future of Indoor Localization systems, Indoor Positioning Systems, and Location-Based Services; the paper presents a comprehensive comparative study of different machine learning approaches that include—Random Forest, Artificial Neural Network, Decision Tree, Support Vector Machine, k-NN, Gradient Boosted Trees, Deep Learning, and Linear Regression. The performance characteristics of each of these learning methods were studied by evaluating the RMSE in X-direction, the RMSE in Y-direction, and the Horizontal Error as per ISO/IEC18305:2016 [31]. The results and findings of this study show that the Random Forest approach can be considered as the optimal learning method for development of such technologies for all practical purposes.

To the best knowledge of the authors, no similar work has been done yet and no work in the field of Indoor Localization thus far has achieved such a superior performance accuracy (RMSE for detection of X coordinate: 5.85 cm, RMSE for detection of Y coordinate: 5.36 cm, and Horizontal Error: 7.93 cm) as presented in this work. Future work would involve—(1) Implementing and deploying all these proposed approaches for Indoor Localization in real-time in different IoT-based environments by using the Context-Driven Human Activity Recognition Framework [64]. For real-time implementation of all these proposed approaches, we plan on conducting experiments as per Institutional Review Board (IRB) approved protocols by setting up an experiment procedure for data collection and analysis. The specific functionalities and characteristic features of the different methodologies that we have outlined in Section 4.1, Section 4.2, Section 4.3 would then be implemented in real-time. Thereafter, the performance characteristics from the real-time data would be studied and compared with the findings presented in Section 5.1, Section 5.2, Section 5.3 (2) Extending the functionalities of the two 'zone'-based Indoor Localization approaches and evaluating their performance characteristics by using the RMSE approach as well as by using some of the other performance metrics defined in ISO/IEC18305:2016 [31]. This would be performed either by analyzing the real-time data collected from (1) or by using a different dataset that consists of user interaction data related to different ADLs and the spatial coordinates of the user's varying position recorded during the dynamic user interactions associated with these different activities.

**Author Contributions:** Conceptualization, N.T. and C.Y.H.; methodology, N.T.; software, N.T.; validation, N.T.; formal analysis, N.T.; investigation, N.T.; resources, N.T.; data curation, N.T.; visualization, N.T.; data analysis and results, N.T.; writing—original draft preparation, N.T.; writing—review and editing, N.T. and C.Y.H.; supervision, C.Y.H.; project administration, C.Y.H.; funding acquisition, Not Applicable. All authors have read and agreed to the published version of the manuscript.

**Funding:** This research received no external funding.

## References

1. Langlois, C.; Tiku, S.; Pasricha, S. Indoor Localization with Smartphones: Harnessing the Sensor Suite in Your Pocket. *IEEE Consum. Electron. Mag.* **2017**, *6*, 70–80. [CrossRef]
2. Zafari, F.; Papapanagiotou, I.; Devetsikiotis, M.; Hacker, T.J. An iBeacon based proximity and indoor localization system. *arXiv* **2017**, arXiv:1703.07876.
3. Dardari, D.; Closas, P.; Djuric, P.M. Indoor Tracking: Theory, Methods, and Technologies. *IEEE Trans. Veh. Technol.* **2015**, *64*, 1263–1278. [CrossRef]
4. Thakur, N. Framework for a Context Aware Adaptive Intelligent Assistant for Activities of Daily Living. Master's Thesis, University of Cincinnati, Cincinnati, OH, USA, 2019. Available online: http://rave.ohiolink.edu/etdc/view?acc_num=ucin15535 28536685873 (accessed on 10 December 2020).
5. Thakur, N.; Han, C.Y. An Improved Approach for Complex Activity Recognition in Smart Homes. Available online: https://link.springer.com/chapter/10.1007/978-3-030-22888-0_15 (accessed on 3 March 2021).
6. United Nations: 2020 Report on Ageing. Available online: http://www.un.org/en/sections/issuesdepth/ageing/ (accessed on 22 November 2020).
7. United States Census Bureau Report: An Aging World: 2015. Available online: https://www.census.gov/library/publications/2016/demo/P95-16-1.html (accessed on 17 December 2020).
8. Key Facts of Dementia. Available online: https://www.who.int/news-room/fact-sheets/detail/dementia (accessed on 24 November 2020).
9. Musa, A.; Nugraha, G.D.; Han, H.; Choi, D.; Seo, S.; Kim, J. A decision tree-based NLOS detection method for the UWB indoor location tracking accuracy improvement. *Int. J. Commun. Syst.* **2019**, *32*, e3997. [CrossRef]
10. Yim, J. Introducing a decision tree-based indoor positioning technique. *Expert Syst. Appl.* **2008**, *34*, 1296–1302. [CrossRef]

11.  Sjoberg, M.; Koskela, M.; Viitaniemi, V.; Laaksonen, J. Indoor location recognition using fusion of SVM-based visual classifiers. In Proceedings of the 2010 IEEE International Workshop on Machine Learning for Signal Processing, Kittila, Finland, 29 August–1 September 2010; pp. 343–348. [CrossRef]

12.  Zhang, S.; Guo, J.; Wang, W.; Hu, J. Indoor 2.5D Positioning of WiFi Based on SVM. In Proceedings of the 2018 Ubiquitous Positioning, Indoor Navigation and Location-Based Services (UPINLBS) Conference, Wuhan, China, 22–23 March 2018; pp. 1–7. [CrossRef]

13.  Zhang, L.; Zhao, C.; Wang, Y.; Dai, L. Fingerprint-based Indoor Localization using Weighted K-Nearest Neighbor and Weighted Signal Intensity. In Proceedings of the 2nd International Conference on Artificial Intelligence and Advanced Manufacture, Association for Computing Machinery (ACM), Manchester, UK, 15–17 October 2020; pp. 185–191.

14.  Ge, X.; Qu, Z. Optimization WIFI indoor positioning KNN algorithm location-based fingerprint. In Proceedings of the 2016 7th IEEE International Conference on Software Engineering and Service Science (ICSESS), Beijing, China, 26–28 August 2016; pp. 135–137.

15.  Hu, J.; Liu, D.; Yan, Z.; Liu, H. Experimental Analysis on Weight KK -Nearest Neighbor Indoor Fingerprint Positioning. *IEEE Internet Things J.* **2018**, *6*, 891–897. [CrossRef]

16.  Khan, I.U.; Ali, T.; Farid, Z.; Scavino, E.; Rahman, M.A.A.; Hamdi, M.; Qiao, G. An improved hybrid indoor positioning system based on surface tessellation artificial neural network. *Meas. Control.* **2020**, *53*, 1968–1977. [CrossRef]

17.  Labinghisa, B.A.; Lee, D.M. Neural network-based indoor localization system with enhanced virtual access points. *J. Supercomput.* **2021**, *77*, 638–651. [CrossRef]

18.  Qin, F.; Zuo, T.; Wang, X. CCpos: WiFi Fingerprint Indoor Positioning System Based on CDAE-CNN. *Sensors* **2021**, *21*, 1114. [CrossRef]

19.  Varma, P.S.; Anand, V. Random Forest Learning Based Indoor Localization as an IoT Service for Smart Buildings. *Wirel. Pers. Commun.* **2020**, 1–19. [CrossRef]

20.  Gao, J.; Li, X.; Ding, Y.; Su, Q.; Liu, Z. WiFi-Based Indoor Positioning by Random Forest and Adjusted Cosine Similarity. In Proceedings of the 2020 Chinese Control and Decision Conference (CCDC), Hefei, China, 22–24 August 2020; pp. 1426–1431.

21.  Ben Jamâa, M.; Koubâa, A.; Baccour, N.; Kayani, Y.; Al-Shalfan, K.; Jmaiel, M. EasyLoc: Plug-and-Play RSS-Based Localization in Wireless Sensor Networks. In *Complex Networks & Their Applications IX*; Springer: Berlin, Germany, 2013; Volume 507, pp. 77–98.

22.  Barsocchi, P.; Lenzi, S.; Chessa, S.; Furfari, F. Automatic virtual calibration of range-based indoor localization systems. *Wirel. Commun. Mob. Comput.* **2011**, *12*, 1546–1557. [CrossRef]

23.  Zhang, Q.; Wang, Y. A 3D mobile positioning method based on deep learning for hospital applications. *EURASIP J. Wirel. Commun. Netw.* **2020**, 1–15. [CrossRef]

24.  Poulose, A.; Han, D.S. Hybrid Deep Learning Model Based Indoor Positioning Using Wi-Fi RSSI Heat Maps for Autonomous Applications. *Electronics* **2020**, *10*, 2. [CrossRef]

25.  Wang, Y.; Lei, Y.; Zhang, Y.; Yao, L. A robust indoor localization method with calibration strategy based on joint distribution adaptation. *Wirel. Netw.* **2021**, 1–15. [CrossRef]

26.  Lin, Y.-T.; Yang, Y.-H.; Fang, S.-H. A case study of indoor positioning in an unmodified factory environment. In Proceedings of the 2014 International Conference on Indoor Positioning and Indoor Navigation (IPIN), Busan, Korea, 27–30 October 2014; pp. 721–722.

27.  Liu, J.; Chen, R.; Chen, Y.; Pei, L.; Chen, L. iParking: An Intelligent Indoor Location-Based Smartphone Parking Service. *Sensors* **2012**, *12*, 14612–14629. [CrossRef]

28.  Jiang, L.; Hoe, L.N.; Loon, L.L. Integrated UWB and GPS location sensing system in hospital environment. In Proceedings of the 2010 5th IEEE Conference on Industrial Electronics and Applications, Taichung, Taiwan, 15–17 June 2010; pp. 286–289.

29.  Barral, V.; Suárez-Casal, P.; Escudero, C.J.; García-Naya, J.A. Multi-Sensor Accurate Forklift Location and Tracking Simulation in Industrial Indoor Environments. *Electronics* **2019**, *8*, 1152. [CrossRef]

30.  Zadeh, A.M.H.; Koo, A.C.; Abadi, H.G.N. Design of Students' Attendance system based on mobile indoor location. In Proceedings of the International Conference on Mobile Learning, Application & Services, Malacca, Malaysia, 18–20 September 2012; pp. 1–4.

31.  ISO/IEC 18305:2016 Information Technology—Real Time Locating Systems—Test and Evaluation of Localization and Tracking Systems. Available online: https://www.iso.org/standard/62090.html (accessed on 13 February 2021).

32.  EVARILOS—Evaluation of RF-based Indoor Localization Solutions for the Future Internet. Available online: https://www2.tkn.tu-berlin.de/tkn-projects/evarilos/index.php (accessed on 13 February 2021).

33.  Correa, A.; Llado, M.B.; Morell, A.; Vicario, J.L.; Barcelo, M. Indoor Pedestrian Tracking by On-Body Multiple Receivers. *IEEE Sens. J.* **2016**, *16*, 2545–2553. [CrossRef]

34.  Bolic, M.; Rostamian, M.; Djuric, P.M. Proximity Detection with RFID: A Step toward the Internet of Things. *IEEE Pervasive Comput.* **2015**, *14*, 70–76. [CrossRef]

35.  Angermann, M.; Robertson, P. FootSLAM: Pedestrian Simultaneous Localization and Mapping without Exteroceptive Sensors—Hitchhiking on Human Perception and Cognition. *Proc. IEEE* **2012**, *100*, 1840–1848. [CrossRef]

36.  Evennou, F.; Marx, F. Advanced Integration of WiFi and Inertial Navigation Systems for Indoor Mobile Positioning. *EURASIP J. Adv. Signal Process.* **2006**, *2006*, 86706. [CrossRef]

37. Wang, H.; Lenz, H.; Szabo, A.; Bamberger, J.; Hanebeck, U.D. WLAN-Based Pedestrian Tracking Using Particle Filters and Low-Cost MEMS Sensors. In Proceedings of the 2007 4th Workshop on Positioning, Navigation and Communication, Hannover, Germany, 22 March 2007; pp. 1–7.

38. Klingbeil, L.; Wark, T. A Wireless Sensor Network for Real-Time Indoor Localisation and Motion Monitoring. In Proceedings of the 2008 International Conference on Information Processing in Sensor Networks (ipsn 2008), St. Louis, MO, USA, 22–24 April 2008; pp. 39–50.

39. Pei, L.; Liu, J.; Guinness, R.; Chen, Y.; Kuusniemi, H.; Chen, R. Using LS-SVM Based Motion Recognition for Smartphone Indoor Wireless Positioning. *Sensors* **2012**, *12*, 6155–6175. [CrossRef]

40. Liu, J.; Chen, R.; Pei, L.; Guinness, R.; Kuusniemi, H. A Hybrid Smartphone Indoor Positioning Solution for Mobile LBS. *Sensors* **2012**, *12*, 17208–17233. [CrossRef] [PubMed]

41. Chen, Z.; Zou, H.; Jiang, H.; Zhu, Q.; Soh, Y.C.; Xie, L. Fusion of WiFi, Smartphone Sensors and Landmarks Using the Kalman Filter for Indoor Localization. *Sensors* **2015**, *15*, 715–732. [CrossRef] [PubMed]

42. Li, Y.; Zhang, P.; Lan, H.; Zhuang, Y.; Niu, X.; El-Sheimy, N. A modularized real-time indoor navigation algorithm on smartphones. In Proceedings of the 2015 International Conference on Indoor Positioning and Indoor Navigation (IPIN), Banff, AB, Canada, 13–16 October 2015; pp. 1–7.

43. Chen, Z.; Zhu, Q.; Soh, Y.C. Smartphone Inertial Sensor-Based Indoor Localization and Tracking With iBeacon Corrections. *IEEE Trans. Ind. Inform.* **2016**, *12*, 1540–1549. [CrossRef]

44. Statistics: Average Size of Newly Built One-Bedroom Apartments in the United States from 2008 to 2018. Available online: https://www.statista.com/statistics/943956/size-newly-built-one-bed-apartments-usa/ (accessed on 7 February 2021).

45. Statistics: Average Size of Newly Built Two-Bedroom Apartments in the United States from 2008 to 2018. Available online: https://www.statista.com/statistics/943958/size-newly-built-two-bed-apartments-usa/ (accessed on 7 February 2021).

46. Ranieri, C.; MacLeod, S.; Dragone, M.; Vargas, P.; Romero, R.A. Activity Recognition for Ambient Assisted Living with Videos, Inertial Units and Ambient Sensors. *Sensors* **2021**, *21*, 768. [CrossRef] [PubMed]

47. Fahada, L.G.; Tahir, S.F. Activity recognition and anomaly detection in smart homes. *J. Neurocomput.* **2021**, *423*, 362–372. [CrossRef]

48. Suriani, N.S.; Rashid, F.A.N. Smartphone Sensor Accelerometer Data for Human Activity Recognition Using Spiking Neural Network. *Int. J. Machine Learn. Comput.* **2021**, *11*, 298–303.

49. Mousavi, S.A.; Heidari, F.; Tahami, E.; Azarnoosh, M. Fall detection system via smart phone and send people location. In Proceedings of the 2020 28th European Signal Processing Conference (EUSIPCO), Amsterdam, The Netherlands, 18–22 January 2021; pp. 1605–1607.

50. Alarifi, A.; Alwadain, A. Killer heuristic optimized convolution neural network-based fall detection with wearable IoT sensor devices. *J. Meas.* **2021**, *167*, 108258. [CrossRef]

51. Al-Okby, M.F.R.; Al-Barrak, S.S. New Approach for Fall Detection System Using Embedded Technology. In Proceedings of the 24th IEEE International Conference on Intelligent Engineering Systems (INES), Reykjavík, Iceland, 8–10 July 2020; pp. 209–214.

52. Nikoloudakis, Y.; Panagiotakis, S.; Markakis, E.; Pallis, E.; Mastorakis, G.; Mavromoustakis, C.X.; Dobre, C. A Fog-Based Emergency System for Smart Enhanced Living Environments. *IEEE Cloud Comput.* **2016**, *3*, 54–62. [CrossRef]

53. Navarro, J.; Vidaña-Vila, E.; Alsina-Pagès, R.M.; Hervás, M. Real-Time Distributed Architecture for Remote Acoustic Elderly Monitoring in Residential-Scale Ambient Assisted Living Scenarios. *Sensors* **2018**, *18*, 2492. [CrossRef]

54. Nikoloudakis, Y.; Markakis, E.; Mastorakis, G.; Pallis, E.; Skianis, C. An NF V-powered emergency system for smart enhanced living environments. In Proceedings of the 2017 IEEE Conference on Network Function Virtualization and Software Defined Networks (NFV-SDN), Berlin, Germany, 6–8 November 2017; pp. 258–263.

55. Facchinetti, D.; Psaila, G.; Scandurra, P. Mobile cloud computing for indoor emergency response: The IPSOS assistant case study. *J. Reliab. Intell. Environ.* **2019**, *5*, 173–191. [CrossRef]

56. Fundació Ave Maria, A Non-Profit Organization in Spain. Available online: https://www.avemariafundacio.org/ (accessed on 3 March 2021).

57. Anderson, G.O. Technology Use and Attitude among Mid-Life and Older Americans. *AARP Res.* **2018**, 1–29. [CrossRef]

58. Vaportzis, E.; Clausen, M.G.; Gow, A.J. Older Adults Perceptions of Technology and Barriers to Interacting with Tablet Computers: A Focus Group Study. *Front. Psychol.* **2017**, *8*. [CrossRef]

59. Elguera Paez, L.; Zapata Del Río, C. Elderly Users and Their Main Challenges Usability with Mobile Applications: A Systematic Review. In *Design, User Experience, and Usability. Design Philosophy and Theory*; HCII 2019, Lecture Notes in Computer Science; Marcus, A., Wang, W., Eds.; Springer: Berlin, Germany, 2019; Volume 11583, pp. 423–438.

60. Mierswa, I.; Wurst, M.; Klinkenberg, R.; Scholz, M.; Euler, T. YALE: Rapid prototyping for complex data mining tasks. In Proceedings of the 12th ACM SIGKDD International Conference on Knowledge Discovery and Data Mining (KDD '06), Philadelphia, PA, USA, 20–23 August 2006; pp. 935–940.

61. Waikato Environment for Knowledge Analysis (WEKA). Available online: https://en.wikipedia.org/wiki/Weka_(machine_learning) (accessed on 16 February 2021).

62. MLC++. Available online: http://robotics.stanford.edu/~{}ronnyk/mlc.html (accessed on 16 February 2021).

63. Saguna, S.; Zaslavsky, A.; Chakraborty, D. Complex activity recognition using context-driven activity theory and activity signatures. *ACM Trans. Comput. Interact.* **2013**, *20*, 1–34. [CrossRef]

64. Chakraborty, S.; Han, C.Y.; Zhou, X.; Wee, W.G. A Context Driven Human Activity Recognition Framework. In Proceedings of the 2016 International Conference on Health Informatics and Medical Systems, Monte Carlo Resort, Las Vegas, NV, USA, 25–28 July 2016; pp. 96–102.
65. Tao, M. A Framework for Modeling and Capturing Social Interactions. Ph.D. Thesis, University of Cincinnati, Cincinnati, OH, USA, 2014. Available online: https://etd.ohiolink.edu/apexprod/rws_olink/r/1501/10?clear=10&p10_accession_num=ucin1423581254 (accessed on 12 February 2021).
66. Tabbakha, N.E.; Ooi, C.P.; Tan, W.H. A Dataset for Elderly Action Recognition Using Indoor Location and Activity Tracking Data. *Mendeley Data*. 2020. [CrossRef]
67. Indoor Positioning Dataset of Bluetooth Beacons Readings Indoor. Available online: https://www.kaggle.com/liwste/indoor-positioning (accessed on 29 December 2020).
68. Root Mean Square. Available online: https://en.wikipedia.org/wiki/Root_mean_square (accessed on 13 February 2021).
69. Confusion Matrix. Available online: https://en.wikipedia.org/wiki/Confusion_matrix (accessed on 13 February 2021).
70. Taipei 101. Available online: https://en.wikipedia.org/wiki/Taipei_101 (accessed on 9 February 2021).
71. Burj Khalifa. Available online: https://en.wikipedia.org/wiki/Burj_Khalifa (accessed on 9 February 2021).
72. Masud, T.; Morris, R.O. Epidemiology of falls. *Age Ageing* **2001**, *30*, 3–7. [CrossRef]